\documentclass[]{aastex631}

\def\figref#1{Figure~\ref{#1}} %\def\figref#1{\hbox{Figure~\ref{#1}}} 
\def\tabref#1{Table~\ref{#1}} %\def\tabref#1{\hbox{Table~\ref{#1}}} 
\def\eqref#1{Equation~\ref{#1}} %\def\eqref#1{\hbox{Eq.~(\ref{#1})}} 
\def\secref#1{Section~\ref{#1}} %\def\secref#1{\hbox{Section~\ref{#1}}} 

\def\comment#1{\textcolor{cyan}{\bf [NT] #1}} %
\def\revise#1{\textcolor{red}{\bf #1}}

\def\rxj{RX J1713.7$-$3946}

\def\hessj1731{HESS J1731$-$347}

\def\bat{{\it Swift}-BAT}

\def\integral{{\it INTEGRAL}}

\def\spi{{\it INTEGRAL}-SPI}

\def\comptel{COMPTEL}

\def\fermi{{\it Fermi}}
\def\lat{{\it Fermi}-LAT}

\def\ams{AMS-02}
\def\voyager{\textit{Voyager}}

\def\galprop{{\sc GalProp}}

\def\flux{\hbox{${\rm erg}~{\rm cm}^{-2}~{\rm s}^{-1}$}}
\def\Mflux{\hbox{MeV~cm$^{-2}$~s$^{-1}$~sr$^{-1}$}}

\def\pizero{\hbox{$\pi^0$}}

\usepackage[]{acro}

\DeclareAcronym{gde}{
  short = GDE ,
  long  = Galactic Diffuse Emission,
}

\DeclareAcronym{cgb}{
  short = CGB ,
  long  = cosmic gamma-ray background,
}

\DeclareAcronym{grb}{
  short = GRB ,
  long  = gamma-ray burst,
}
 
\DeclareAcronym{agn}{
  short = AGN ,
  long  = active galactic nucleus ,
  long-plural-form = active galactic nuclei
}
\DeclareAcronym{fsrq}{
  short = FSRQ ,
  long  = flat-spectrum radio quasar,
}
  
\DeclareAcronym{lmxb}{
  short = LMXB ,
  long  = low mass X-ray binary ,
  long-plural-form = low mass X-ray binaries ,
}
\DeclareAcronym{hmxb}{
  short = HMXB ,
  long  = high mass X-ray binary ,
  long-plural-form = high mass X-ray binaries ,
}
  
\DeclareAcronym{ic}{
  short = IC ,
  long  = inverse Compton ,
}

\DeclareAcronym{nustar}{
  short = {\it NuSTAR} ,
  long  = Nuclear Spectroscopic Telescope Array ,
}

\DeclareAcronym{bi}{
  short = BI ,
  long  = backside illumination ,
}
\DeclareAcronym{fi}{
  short = FI ,
  long  = frontside illumination ,
}
\DeclareAcronym{fov}{
  short = FoV ,
  long  = field of view ,
}
  
\DeclareAcronym{sim}{
  short = SIM ,
  long  = Science Instrument Module ,
}
\DeclareAcronym{hetg}{
  short = HETG ,
  long  = High Energy Transmission Grating ,
}
\DeclareAcronym{letg}{
  short = LETG ,
  long  = Low Energy Transmission Grating ,
}
\DeclareAcronym{hrc}{
  short = HRC ,
  long  = High Resolution Camera ,
}
\DeclareAcronym{acis}{
  short = ACIS ,
  long  = Advanced CCD Imaging Spectrometer ,
}
\DeclareAcronym{hrma}{
  short = HRMA ,
  long  = High Resolution Mirror Assembly,
}

\DeclareAcronym{compton}{
  short = Compton ,
  long  = Compton Gamma Ray Observatory,
}
\DeclareAcronym{hst}{
  short = HST ,
  long  = Hubble Space Telescope ,
}
\DeclareAcronym{iact}{
  short = IACT ,
  long  = Imaging Atmospheric Cherenkov Telescope ,
}
\DeclareAcronym{wcd}{
  short = WCD ,
  long  = Water Cherenkov Detector ,
}

\DeclareAcronym{hawc}{
  short = HAWC ,
  long  = High-Altitude Water Cherenkov ,
}
\DeclareAcronym{cta}{
  short = CTA ,
  long  = Cherenkov Telescope Array ,
}

\DeclareAcronym{em}{
  short = EM ,
  long  = electromagnetic ,
}
\DeclareAcronym{ism}{
  short = ISM ,
  long  = interstellar medium ,
}
\DeclareAcronym{csm}{
  short = CSM ,
  long  = circumstellar medium ,
}
\DeclareAcronym{sne}{
  short = SNe ,
  long  = supernovae , 
}

\DeclareAcronym{iss}{
  short = ISS ,
  long  = International Space Station ,
}

\DeclareAcronym{uhecr}{
  short = UHECRs ,
  long  = ultra high energy cosmic rays , 
}
\DeclareAcronym{ta}{
  short = TA ,
  long  = Telescope Array , 
}
\DeclareAcronym{auger}{
  short = Auger ,
  long  = Pierre Auger Observatory , 
}
\DeclareAcronym{ams}{
  short = AMS ,
  long  = Alpha Magnetic Spectrometer , 
}
\DeclareAcronym{pamela}{
  short = PAMELA ,
  long  = Payload for Antimatter Matter Exploration and Light-nuclei Astrophysics , 
}
   
\DeclareAcronym{cmb}{
  short = CMB ,
  long  = Cosmic Microwave Background , 
}
\DeclareAcronym{sed}{
  short = SED ,
  long  = spectral energy distribution , 
}
\DeclareAcronym{mhd}{
  short = MHD ,
  long  = magnetohydrodynamical ,
}
\DeclareAcronym{dof}{
  short = dof ,
  long  = degree of freedom ,
}
\DeclareAcronym{cco}{
  short = CCO ,
  long  = central compact object ,
  first-style = default
}
\DeclareAcronym{lmc}{
  short = LMC ,
  long  = Large Magellanic Cloud ,
}
\DeclareAcronym{smc}{
  short = SMC ,
  long  = Small Magellanic Cloud ,
}

\DeclareAcronym{hess}{
  short = H.E.S.S. ,
  long  = High Energy Spectroscopic System ,
  first-style = default
}
\DeclareAcronym{snr}{
  short = SNR ,
  long  = supernova remnant ,
}
\DeclareAcronym{pwn}{
  short = PWN ,
  short-plural = e ,
  long  = pulsar wind nebula ,
  long-plural  = e ,
}

\DeclareAcronym{sn}{
  short = SN ,
  short-plural = e ,
  long  = supernova ,
  long-plural  = e ,
  first-style = default
}

\DeclareAcronym{nw}{
  short = NW ,
  long  = northwest ,
  first-style = default
}
\DeclareAcronym{hxc}{
  short = HXC ,
  long  = hard X-ray component ,
  first-style = default
}

\DeclareAcronym{cr}{
  short = CR ,
  long  = cosmic ray ,
}
\DeclareAcronym{psf}{
  short = PSF ,
  long  = point spread function ,
}
\DeclareAcronym{hpd}{
  short = HPD ,
  long  = half power diameter ,
}
\DeclareAcronym{fwhm}{
  short = FWHM ,
  long  = full width of half maximum ,
}

\DeclareAcronym{pic}{
  short = PIC ,
  long  = particle-in-cell ,
  tag = numerical ,
}
\DeclareAcronym{cxb}{
  short = CXB ,
  long  = Cosmic X-ray Background ,
}
\DeclareAcronym{grxe}{
  short = GRXE ,
  long  = Galactic Ridge X-ray Emission ,
}
\DeclareAcronym{pa}{
  short = PA ,
  long  = Positional Angle ,
}
\DeclareAcronym{dsa}{
  short = DSA ,
  long  = diffusive shock acceleration ,
}

\def\pizero{\hbox{$\pi^0$}}

\def\degr{\hbox{$^\circ$}}

\def\utw{\smash{\rlap{\lower5pt\hbox{$\sim$}}}}
\def\udtw{\smash{\rlap{\lower6pt\hbox{$\approx$}}}}

\def\min{\hbox{$^{\rm m}$}}

\def\apjl{ApJ}%
\def\apjs{ApJS}%
\def\ao{Appl.~Opt.}%
\def\aap{A\&A}%
\received{\today}
\revised{MM DD, YYYY}
\accepted{MM DD, YYYY}

\submitjournal{ApJ}

\shorttitle{
Source contribution to MeV gamma-ray diffuse emission
}
\shortauthors{Tsuji et al.}
\graphicspath{{./}{figures/}}
\begin{document}

\title{
%The origin of MeV Gamma-Ray Diffuse Emission from the Inner Galactic region
MeV Gamma-Ray Source Contribution to the Inner Galactic Diffuse Emission
}

%% you desire. In v6.31 you can include a footnote in the title.

%\correspondingauthor{August Muench}
%\email{greg.schwarz@aas.org, gus.muench@aas.org}

\correspondingauthor{Naomi Tsuji}
%\email{n.tsuji@rikkyo.ac.jp}
%\email{naomi.tsuji@riken.jp}
\email{ntsuji@kanagawa-u.ac.jp}

\author[0000-0002-0786-7307]{Naomi Tsuji}
\affiliation{Faculty of Science, Kanagawa University, 2946 Tsuchiya, Hiratsuka-shi, Kanagawa 259-1293, Japan}
\affiliation{Interdisciplinary Theoretical \& Mathematical Science Program (iTHEMS), RIKEN, 2-1 Hirosawa, Wako, Saitama 351-0198, Japan}
\affiliation{Department of Physics, Rikkyo University, 3-34-1 Nishi Ikebukuro, Toshima-ku, Tokyo 171-8501, Japan}

\author[0000-0002-7272-1136]{Yoshiyuki Inoue}
\affiliation{Department of Earth and Space Science, Graduate School of Science, Osaka University, Toyonaka, Osaka 560-0043, Japan}
\affiliation{Interdisciplinary Theoretical \& Mathematical Science Program (iTHEMS), RIKEN, 2-1 Hirosawa, Wako, Saitama 351-0198, Japan}
\affiliation{Kavli Institute for the Physics and Mathematics of the Universe (WPI), The University of Tokyo, Kashiwa 277-8583, Japan}

\author{Hiroki Yoneda}
\affiliation{Nishina Center, RIKEN, 2-1 Hirosawa, Wako, Saitama 351-0198, Japan}

%\author{Tsuguo Aramaki}
%\affiliation{Northeastern University, 360 Huntington Ave, Boston, MA 02115, USA}
%\affiliation{SLAC National Accelerator Laboratory/Kavli Institute for Particle Astrophysics and Cosmology, Menlo Park, California 94025, USA}

%\author{Georgia Karagiorgi}
%\affiliation{Columbia University, New York, NY, 10027, USA}

\author{Reshmi Mukherjee}
%\affiliation{Columbia University, New York, NY, 10027, USA}
\affiliation{Department of Physics and Astronomy, Barnard College, Columbia University, New York, NY, 10027, USA}
%\affiliation{Department of Physics, Columbia University, 538 West 120th Street, New York, NY 10027, USA}
%Department of Physics and Astronomy, Barnard College, Columbia University, New York, USA}

\author{Hirokazu Odaka}
\affiliation{Department of Physics, The University of Tokyo, 7-3-1 Hongo, Bunkyo, Tokyo 113-0033, Japan}
\affiliation{Kavli Institute for the Physics and Mathematics of the Universe (WPI), The University of Tokyo, Kashiwa 277-8583, Japan}

%% automatically takes care of all commas and “and”s between authors names.


%% Mark off the abstract in the ``abstract’’ environment.
\begin{abstract}
%%%%%%%%%%%%%%%%%%%%%%%%%
%%%%%%%%%%%%%%%%%%%%%%%%%

The origin of the inner Galactic emission, measured by \comptel\ with a flux of $\sim 10^{-2}$~\Mflux\ in the 1–30 MeV energy range
%\revise{\sout{from a region of $|\ell | \leq 60\degr$ and $|b| \leq 10\degr$}},
from the inner Galactic region,
has remained unsettled since its discovery.
%In this paper, we investigate the origin of this emission by taking into account the Galactic diffuse emission and individual MeV gamma-ray sources which were not resolved by \comptel. The Galactic diffuse emission is calculated by \galprop\ to reconcile the cosmic-ray and gamma-ray spectra with observations by \ams, \voyager, and \lat, resulting in a flux of (2–8)$\times 10^{-3}$~\Mflux. The source contribution is estimated for sources crossmatched between the \bat\ and \lat\ catalogs by interpolating the energy spectra in the hard X-ray and GeV gamma-ray ranges, resulting in a flux of at least 10\% of the \comptel\ excess.
%estimated by extrapolating the energy spectra in the hard X-ray and GeV gamma-ray ranges, which are respectively observed by \bat\ and \lat. We find that the source contribution is at least $\sim$10\% of the \comptel\ emission.
%Although the Galactic diffuse emission is uncertain by a factor of few, the inner Galactic emission could be roughly reproduced by the Galactic diffuse emission and the MeV gamma-ray sources.
In this paper,
we elaborate on a model of individual MeV gamma-ray sources unresolved by \comptel.
This is conducted for sources crossmatched between the \bat\ and \lat\ catalogs by interpolating the energy spectra in the hard X-ray and GeV gamma-ray ranges,
as well as unmatched sources between the two catalogs.
We find that the source contribution to the \comptel\ emission would be at least $\sim$20\%.
Combined with the Galactic diffuse emission, which is not well constrained,
%is calculated by \galprop\ to reconcile the cosmic-ray and gamma-ray spectra with observations by \ams, \voyager, and \lat, resulting in a flux of (2–8)$\times 10^{-3}$~\Mflux.
%\sout{and extragalactic gamma-ray background}?,
the \comptel emission can be roughly reproduced in some cases. %models of Galactic diffuse emission.

%\comment{Abstract should be less than 250 words}

%This example manuscript is intended to serve as a tutorial and template for authors to use when writing their own AAS Journal articles. The manuscript includes a history of \aastex\ and documents the new features in the previous versions as well as the bug fixes in version 6.31. This manuscript includes many figure and table examples to illustrate these new features.  Information on features not explicitly mentioned in the article can be viewed in the manuscript comments or more extensive online documentation. Authors are welcome replace the text, tables, figures, and bibliography with their own and submit the resulting manuscript to the AAS Journals peer review system.  The first lesson in the tutorial is to remind authors that the AAS Journals, the Astrophysical Journal (ApJ), the Astrophysical Journal Letters (ApJL), the Astronomical Journal (AJ), and the Planetary Science Journal (PSJ) all have a 250 word limit for the  abstract\footnote{Abstracts for Research Notes of the American Astronomical  Society (RNAAS) are limited to 150 words}.  If you exceed this length the Editorial office will ask you to shorten it. This abstract has 182 words.

\end{abstract}

%% but this old “keyword” functionality is maintained in case authors want
%(ISM:) cosmic rays ---
%Cosmic rays (329) ---

\section{Introduction}
\label{sec:intro}
%%%%%%%%%%%%%%%%%%%%%%%%%
%%%%%%%%%%%%%%%%%%%%%%%%%

%\comment{ApJL. Less than 3,500 words, not including Acknowledgement, Appendix, and supplementary materials. No more than 5 combined figures (each limited to 9 panels) and tables, e.g. 3 figures and 2 tables. References should be less than 50.
%\textbf{3,419 words on May 25} without Appendix.
%\textbf{3,603 words on Jun 10} without Appendix.
%\url{https://aastex.aas.org/ApJL/countwords.html}.}

%%% Intro
%MeV gap? Feedback on Galaxy evolution? approved COSI?
The MeV gamma-ray domain, in particular $\sim$1--100 MeV, is the only unexplored window among recent multiwavelength observations in astrophysics, often referred to as the ``MeV gap".
The MeV gamma-ray diffuse emission, which has been hinted by the observations of \comptel\ \citep{strong_diffuse_1996}, is one of the unsolved problems in MeV gamma-ray astrophysics.
Investigation of this diffuse emission covers several important facets, including the Galactic diffuse emission from low-energy (sub-GeV) \acp{cr}, individual MeV gamma-ray objects, and/or new populations of MeV gamma-ray radiation originated from dark matter or neutrinos.
Thus, the study of the MeV gamma-ray diffuse emission would have a lot of influence on these broad topics, especially in the next decade when some missions will give us a new insight into the MeV gap.

\begin{figure}[ht!]
\plotone{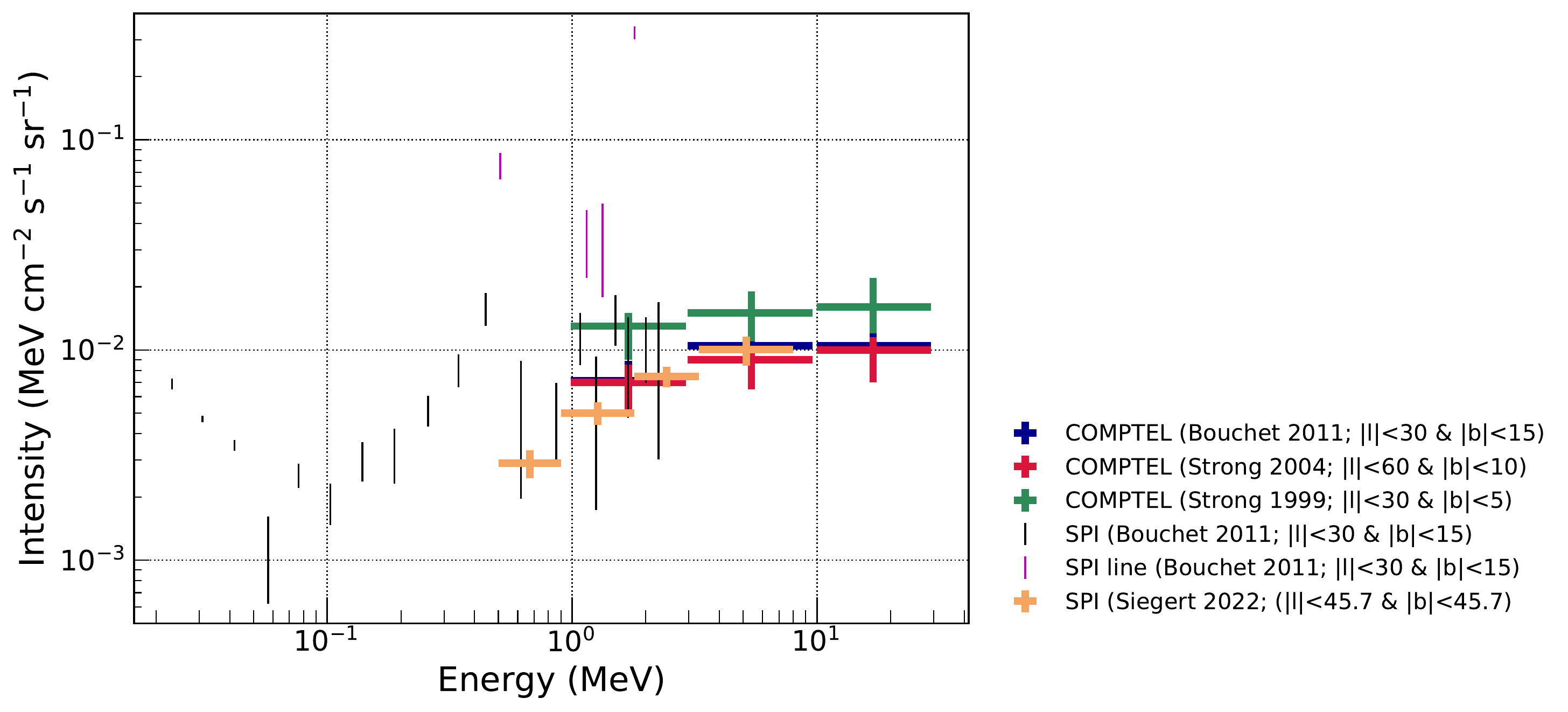}
\caption{
Measurements of the inner Galactic diffuse emission in the MeV gamma-ray range.
%\comment{spell of Bouchet}
\label{fig:size} }
\end{figure}

%%% COMPTEL excess
%\comment{please check this paragraph, @Hiroki.}
The Imaging Compton Telescope \comptel\ onboard the Compton Gamma-Ray Observatory (CGRO) \citep{Comptel} reported the detection of diffuse emission of 10$^{-2}$ \Mflux\ in 1--30 MeV from the inner Galactic region \citep{strong_diffuse_1996}.
This measurement revealed that the emission was actually diffuse, also confirmed by OSSE \citep{strong_diffuse_1996}, and consistent with the result of COS-B \citep{strong_radial_1988}.
The \comptel\ diffuse emission was calculated as follows.
%\comment{clarify what is included in the emission.}
The gamma-ray intensity model consisted of \ac{gde} (i.e., radiation from interactions of \acp{cr} with gas and photon fields), source (only Crab), and an isotropic term.
The instrumental background, which dominated the detected gamma-ray events, was implemented with a fixed spectral shape \citep{strong_diffuse_1994}.
%by fixing the spectral shape and setting the normalization free \citep{strong_diffuse_1994}.
The overall fit was performed by combining the gamma-ray intensity model weighted by the response function and the instrumental background.
Based on the best-fit parameters of \ac{gde}, the aforementioned diffuse emission was estimated.
As shown in \figref{fig:size},
there are several results of the \comptel\ diffuse emission obtained with different models, data (observation phase), and size of the inner Galactic region \citep{strong_diffuse_1994,strong_diffuse_1996,strong_diffuse_2004,bouchet_diffuse_2011},
and the latest one with $|\ell| \leq 30\degr$ and $|b| \leq 15\degr$ \citep{bouchet_diffuse_2011} is used in this paper.
%and the one with $|\ell| \leq 60\degr$ and $|b| \leq 10\degr$ \citep{strong_diffuse_2004} is used in this paper.
%The flux of the diffuse emission was calculated by using a fitting model which consisted of \ac{gde}, source (only Crab), an isotropic term, and the instrumental background. and the instrumental background was subtracted by fixing the spectral shape and setting the normalization free \citep{strong_diffuse_1994}.
%\sout{Since only Crab was taken into account\check{RM: replace this sentence}, the diffuse emission must include radiation from the other MeV gamma-ray sources detected or unresolved by \comptel.

%%% origin of COMPTEL excess
The origin of the inner Galactic diffuse emission by \comptel\ has been in active debate. % for more than 20 years.
Almost two decades since its discovery, an all-sky gamma-ray survey by \lat\ unveiled the detailed spectroscopy of the diffuse emission in the GeV energy band.
It revealed that the diffuse emission observed by \lat\ was fairly explained by a combination of the \ac{gde}, resolved GeV gamma-ray sources, and extragalactic gamma-ray background (i.e., \ac{cgb}) \citep{ackermann_fermi_2012}.
Note that some locally characteristic radiation, such as the \fermi\ bubble and the Galactic center excess, still remain elusive \citep{su_giant_2010,ackermann_fermi_2017,murgia_fermilat_2020}.
If the \ac{gde} model of \lat\ is extrapolated to the MeV energy range,
%there clearly exists an excess component to account for the \comptel\ emission (e.g., \cite{strong_diffuse_2004}), which is referred to as the ``\comptel\ excess".
there is an apparent excess component to account for the \comptel\ emission (e.g., \citealt{strong_diffuse_2004}), which is commonly referred to as the \comptel\ excess.
There are several scenarios for reproducing the \comptel\ excess:
(1) Individual MeV gamma-ray sources should be taken into consideration.
Only the Crab was considered when calculating the \comptel\ emission, although \comptel\ detected 25 steady sources \citep{Schonfelder2000}.
Furthermore, sources which were not resolved by \comptel\ would also have a fraction of the contribution. % to the \comptel\ excess.
(2) There are non-negligible % (more than a factor of few)
uncertainties on the model of \ac{gde}, since it has a lot of unconstrained parameters (e.g., photon field densities, \ac{cr} source distribution, \ac{cr} injection spectra, and propagation mechanism).
Enhancement of one or more of these parameters can make \ac{gde} higher to reach the \comptel\ excess \citep{bouchet_diffuse_2011}.
%(3) \ac{cgb} might not be fully subtracted.
(3) New populations, such as annihilation or decay of dark matter \citep{boddy_indirect_2015,christy_indirect_2022,Binder_2022} and/or cascaded gamma rays accompanying cosmic neutrinos \citep{fang_tev_2022}, might be present.

%%% updates since COMPTEL
We address a few updates on observations of the MeV gamma-ray diffuse emission since \comptel.
\integral-SPI measured a spectrum of the diffuse emission in 0.02–2.4 MeV, which comprised a continuum component and four gamma-ray lines (i.e., positron annihilation, $^{26}$Al and $^{60}$Fe lines) \citep{bouchet_diffuse_2011}.
The continuum component is consistent with the emission by \comptel\ in the overlapping energy range of 1–2.4 MeV, as shown in \figref{fig:size}.
\cite{bouchet_diffuse_2011} argued that the diffuse emission by \integral-SPI can be roughly reproduced by the standard model of \ac{gde}.
The fit, however, became much improved if they increased the normalization of the primary \ac{cr} electron spectrum or the interstellar radiation field in the Galactic bulge or a large Galactic CR halo.
The latest result of the 0.5--8 MeV observation by \integral-SPI was presented in \cite{siegert_diffuse_2022}.
Using the new analysis with the lower level of signal-to-noise ratio, they confirmed that the obtained diffuse emission showed a mismatch of a factor of 2–3 in normalization with respect to the baseline model of \ac{gde}.
The \spi\ spectra are also illustrated in \figref{fig:size}.
%This may arise from enhanced target photon densities and/or electron source spectra, slightly modified diffusion properties, or an unresolved population of MeV gamma-ray sources.
Besides \spi, the electron-tracking Compton camera (ETCC) aboard the balloon mission of SMILE-2$+$ retrieved a gamma-ray lightcurve %\textcolor{blue}{YI: gamma-ray map?}
in 0.15–2.1 MeV during flight, showing enhanced gamma rays when it was pointing at the vicinity of the Galactic center \citep{takada_first_2022}.

%\revise{Add one paragraph about the MeV gamma-ray sources? There are a few detected sources, but the population of undetected/unresolved would also have (negligible) contribution. This is }

%%% ToC
In this paper, we quantitatively estimate how much the unresolved MeV gamma-ray sources would contribute to the \comptel\ excess.
%Combined with \ac{gde}, we investigate the origin of the \comptel\ excess.
\secref{sec:analysis} presents descriptions of the sources and GDE.
The results and discussion are given in \secref{sec:discussion}.
\secref{sec:conclusion} summarizes this study.

\section{%Analysis ---
Components of the inner Galactic diffuse emission }
\label{sec:analysis}
%%%%%%%%%%%%%%%%%%%%%%%%%
%%%%%%%%%%%%%%%%%%%%%%%%%

\subsection{MeV gamma-ray sources}
\label{sec:sources}
%%%%%%%%%%%%%%%%%%%%%%%%%
%%%%%%%%%%%%%%%%%%%%%%%%%

%%% Summary of \bat\ and \lat\ cross-match. (Tsuji+ 2021)
Although the previous studies (e.g., \citealt{strong_diffuse_1996,orlando_imprints_2018,siegert_diffuse_2022}) proposed that the \comptel\ excess would be attributed by radiation from individual unresolved sources,
the quantitative estimation has not been done yet.
%The emission from individual sources also contributes to the diffuse emission.
We estimate this source contribution from a MeV gamma-ray source catalog in \cite{tsuji_cross-match_2021}, which presented a crossmatching between the 105-month \bat\ \citep{Bird2016} and 10-yr \lat\ catalogs \citep{4fgl,4fgldr2}, resulting in 156 point-like and 31 extended crossmatched sources\footnote{
The MeV gamma-ray source catalog is available in \url{https://tsuji703.github.io/MeV-All-Sky}.}.
Note that among them, 136 point sources and 15 extended sources are firmly matched (i.e., the hard X-ray and GeV gamma-ray emission are originated from the same source), and 16 sources were actually detected by \comptel\ \citep{tsuji_cross-match_2021}.
These crossmatched sources, which are both hard X-ray and GeV gamma-ray emitters, are prominent sources in the MeV gamma-ray sky.
%\check{RM: Are these projected sources in MeV gamma rays or have they been actually detected ?}

%%% source breakdown
19 point sources and 14 extended sources in \cite{tsuji_cross-match_2021} are located in the inner Galactic region with $|\ell| \leq 30\degr$ and $|b| \leq 15\degr$.
%The point sources compose 5 blazars, 4 X-ray binaries, 4 pulsars, a Seyfert galaxy (Circinus galaxy), Galactic center (Sgr A$^\star$), a globular cluster (ESO 520$-$27), 4 unidentified sources, and 3 false matches. The extended sources are 8 PNWe, 2 SNRs, 5 spp\footnote{Sources that are candidates of SNR or PNW.}, and 2 unidentified sources.
The point sources compose three blazars, three pulsars, one X-ray binary (RX J1826.2$-$1450), Galactic center (Sgr A$^\star$), one globular cluster (ESO 520$-$27), four unidentified sources, and six false matches\footnote{Falsely matched sources indicate spatially crossmatched sources, but the hard X-ray (\bat) and gamma-ray (\lat) sources are different origins \citep{tsuji_cross-match_2021}.}.
The extended sources are six PNWe, two SNRs, four spp\footnote{Sources that are candidates of SNR or PNW.}, an unidentified source, and a false match.

%Some examples are shown in \figref{fig:sed}
%\textcolor{blue}{YI:Shall we put this figure in the Appendix? Then, we can add some words explaining the fitting procedure.}.

%%% Fitting Model
We jointly fit the \acp{sed} of \bat\ in 14--195 keV and \lat\ in 50 MeV--300 GeV (see also Appendix \ref{sec:sources_appendix}).
%In order to fit the \acp{sed},
We adopt a log-parabola model first.
The log-parabola model can be applied to most of the sources, in particular blazars like \acp{fsrq} or pulsars, of which the MeV gamma-ray radiation is originated from a single electron population.
We find that nine sources (one pulsar, three false-matched sources, one globular cluster, one unidentified, Galactic center, one PWN, and one spp) cannot be well fitted by the log-parabola model, inferred from a large reduced chi-squared of $>$10.
For these sources, we adopt a two-component model, which is a superposition of the models in the \bat\ and \lat\ catalogs.
Additionally, we also use the same two-component model for \rxj, %\check{RM: What sort of 2-component model?}
because it is a \ac{snr}, of which the X-ray and gamma-ray emission do not originate from the same process.
%Examples of 1RXS J174036.3$+$521155 (FSRQ) and Kes 73 (spp) are shown in \figref{fig:sed}.
Likewise, a choice of the fitting model may cause uncertainty on the source flux.
\if0
\comment{move to Appendix? }
The remaining two sources, Circinus galaxy and MSH 15$-$52, are fitted with a source-dependent model as follows. %The hard X-ray emission of Circinus galaxy (a Seyfert galaxy) is reproduced by \textcolor{blue}{\sout{ MyTorus (Gaussian-like) model\footnote{https://www.mytorus.com/}} accretion disk emission}, while its gamma-ray emission is unknown \citep{hayashida_discovery_2013}.
The hard X-ray emission of Circinus galaxy (a Seyfert galaxy) is reproduced by accretion disk emission, while its gamma-ray emission is unknown \citep{hayashida_discovery_2013}.
We apply a combination of a Gaussian model in the hard X-ray band and a power-law model from 4FGL (\figref{fig:sed} in \secref{sec:sources_appendix}).
The emission in the hard X-ray to GeV gamma-ray bands from MSH 15$-$52 could be given by a superposition of emission from the central pulsar and its nebula, which we adopt log-parabola and power-law models, respectively \citep{abdo_detection_2010}.
\comment{move to Appendix?}
\fi
We checked that the choice of the fitting model (i.e., log-parabola or broken power-law models) does not have a large effect
(less than 50\% with respect to the adopted source model shown in Figure 2) on the undermentioned result.

\begin{figure*}[ht!]
%\plotone{figures/SED_mix_grid.pdf}
\plotone{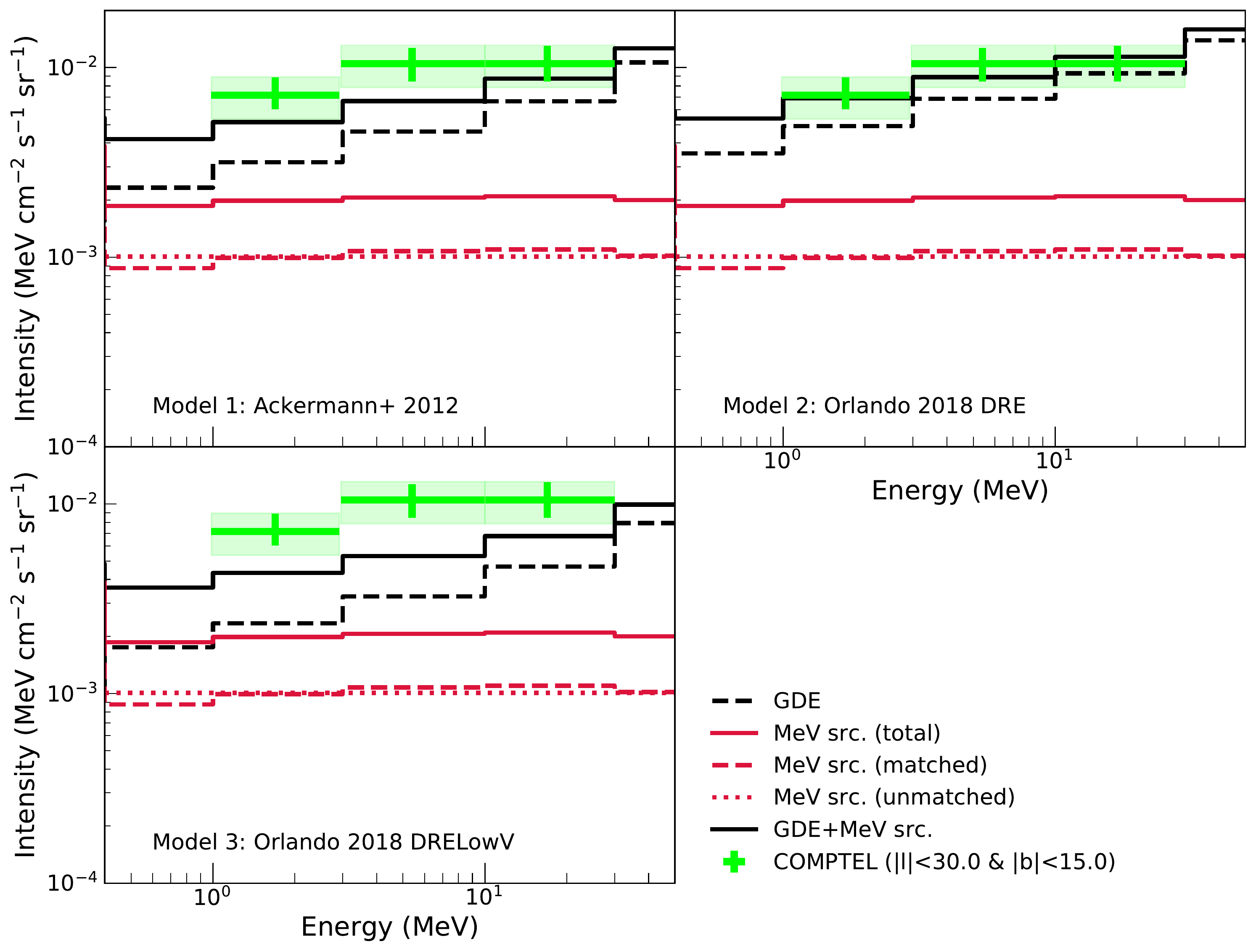}
%\plotone{figures/SED_mix_grid_v1.1_Gal_30.0x15.0_binFalse_CGBfactor0_addUNmatchTrue.pdf}
\caption{
The \acp{sed} of \ac{gde} (dashed black line) and
the sources (solid red line for all the sources, dashed red for the crossmatched sources, and dotted red for the unmatched sources).
The combined spectrum of these two components is illustrated with a solid black line.
The results with \ac{gde} Models 1, 2, and 3 are respectively shown in the upper left, upper right, and lower left.
The \comptel\ emission \citep{bouchet_diffuse_2011} is indicated by light green points with its systematic error \citep{strong_diffuse_1994}.
\label{fig:model}}
\end{figure*}

%%% Contribution to COMPTEL excess
Based on the best-fit model, we estimate spectra in the MeV gamma-ray energy range, sum up all the spectra of the 33 sources in the inner Galactic region, and divide it by the region size.
The result is shown in \figref{fig:model}.
The contribution of all crossmatched sources to the \comptel\ excess is about 10\%,
and the ratio of the point sources to the extended sources is roughly 1–1.5.
It should be noted that if we select only the firmly matched sources in the inner Galactic region (nine point-like and 12 extended sources), the total source contribution is reduced to 60--80\% of the spectrum shown in \figref{fig:model}.
%We also note that contamination from bright sources outside the inner Galactic region (e.g., Cygnus X-1) would make the source spectrum change by a factor of few if included.
%\comment{We also note that contamination from bright sources outside the inner Galactic region (e.g., Cygnus X-1), arising from ring-like response in the imaging analysis of Compton camera, would not largely affect the source spectrum (at most a factor of few).  related to comment \#C4. Cut this sentence??}
%Compared with the GeV gamma-ray source component by \lat\ \citep{ackermann_fermi_2012} in the region with $|\ell| \leq 80\degr$ and $|b| \leq 10\degr$, our source spectrum is roughly in agreement within a factor of $\sim$3 at 200 MeV.

%%% Spectral shape of sources
%Blazars, especially \acp{fsrq} which have their emission peaks around the MeV energy band, would make the combined spectrum convex upward. Indeed, among the 5 blazars inside the inner Galactic region, 2 are \acp{fsrq}, and 3 are bcu\footnote{A source class of a blazar candidate of uncertain type.} (but all FSRQ-like).

%%%%%%%%%%%%%%%%%%%%%%%%%%%%
%%%%%%%%%%%%%%%%%%%%%%%%%%%%

\begin{figure}[ht!]
\plotone{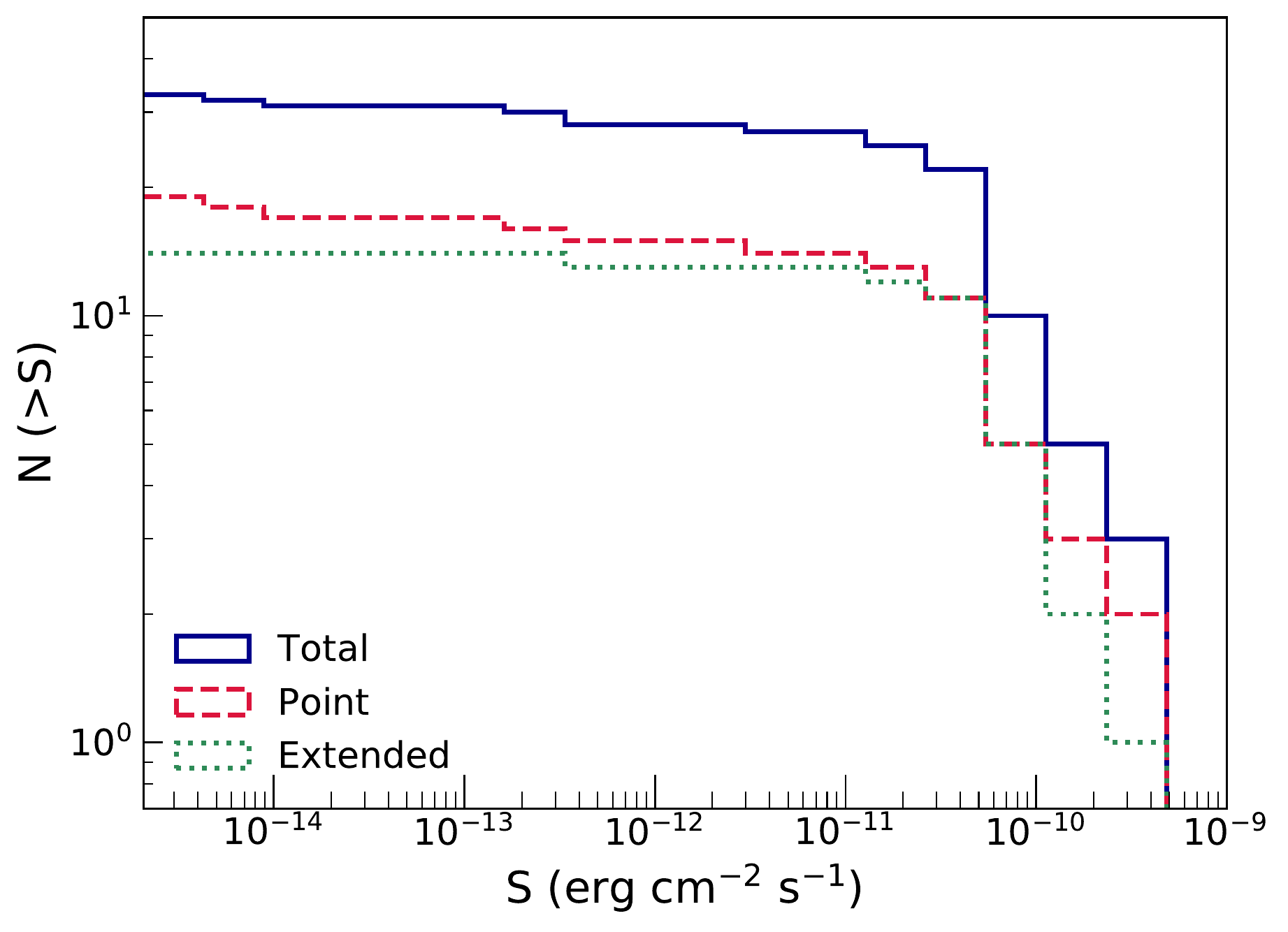}
%\plottwo{figures/Hist_logNlogS_integratedFlux_mix_v2_30x15.pdf}{figures/Hist_logFlogS_integratedFlux_mix_v0_30x15.pdf}
%\plottwo{figures/Hist_logNlogS_integratedFlux_mix_v2.pdf}{figures/Hist_logFlogS_integratedFlux_mix_v0.pdf}
%\plotone{figures/Hist_.pdf}
%\plottwo{figures/Hist_logNlogS.pdf}{figures/Hist_.pdf}
\caption{
%The log N-log S plot (left) and log F-log S plot (right)
The log N-log S plot
of the MeV gamma-ray sources in the inner Galactic region with $|\ell| \leq 30\degr$ and $|b| \leq 15\degr$.
The flux is integrated in 1–10 MeV.
\label{fig:hist}
}
\end{figure}

%%% Log N - log S
\figref{fig:hist} shows a cumulative distribution of the estimated flux of the crossmatched sources in the inner Galactic region (the so-called log N-log S plot).
%\sout{\figref{fig:hist} (right) shows log F-log S plot, where $F$ is given by $F = \int S \frac{dN}{dS} dS$.}
Most of the crossmatched sources have a flux larger than 10$^{-12}$~\flux, and the number of sources decreases towards the higher flux.
In the higher energy part with $S > 10^{-11}$~\flux, the distribution is roughly described by $N(>S) \approx S^{-0.7}$.
It is unknown how this feature is extrapolated or turned over in the lower energy part due to lacking the knowledge of sub-threshold, faint sources, which we discuss in the next paragraph.
%Recalling that the flux sensitivity of the \bat\ and \lat\ catalogs is approximately 10$^{-12}$~\flux, \figref{fig:hist} implies that fainter sources do not have an effect the result since $N(>S) \approx S^{-0.7}$ for $S > 10^{-11}$~\flux.
%\comment{referee comment: we do not know the sub-threshold sources, so this figure cannot state that ``fainter sources do not have an effect the result". The question is how far the power-law of $S^{-0.7}$ can be extrapolated and when does it turn over. The low-flux part ($<$e-12) is unknown.}

%%%%%%%%%%%%%%%%%%%%%%%%%%%%
%%%%%%%%%%%%%%%%%%%%%%%%%%%%

%%% unmatched sources
%%% (1)
Besides the crossmatched sources in \cite{tsuji_cross-match_2021}, there exist many unmatched sources that would have a significant contribution accumulatively.
In the region with $|\ell| \leq 30\degr$ and $|b| \leq 15\degr$, there are 152 \bat\ and 708 \lat\ (4FGL-DR2) sources, where the crossmatched sources are excluded.
We assume that these unmatched sources (860 in total) are fainter than $10^{-12}$ \flux\ in the MeV energy band since they are not detected by \bat\ or \lat, with the sensitivity being approximately 10$^{-12}$~\flux.
Then, the accumulative source flux is $\sim 10^{-3}$ \Mflux, which is roughly comparable to that of the crossmatched sources.
Combined with the crossmatched sources, the contribution of the sources is $\sim$20\% of the \comptel\ excess (\figref{fig:model}).
Compared with the GeV gamma-ray source component by \lat\ \citep{ackermann_fermi_2012} in the region with $|\ell| \leq 80\degr$ and $|b| \leq 10\degr$, our source spectrum is roughly in agreement within a factor of $\sim$2 at 200 MeV.
%The flux of the unmatched sources is also at the same level for the different regions, such as $|\ell| \leq 60\degr$ and $|b| \leq 10\degr$.
The estimation of the accumulated source flux for the different regions, as shown in \figref{fig:size}, is given in Appendix~\ref{sec:region_appendix}.
Detailed modeling of the unmatched sources will be presented in a future publication.

\subsection{Galactic diffuse emission}
\label{sec:gde}
%%%%%%%%%%%%%%%%%%%%%%%%%
%%%%%%%%%%%%%%%%%%%%%%%%%

%%% GALPROP
%We make use of a \ac{cr} propagation code, \galprop\ \revise{(version 54 of WebRun)}, to evaluate \ac{gde}.\galprop\ is designed to calculate astrophysics of \acp{cr} (i.e., propagation and energy loss) and photon emissions in the radio to gamma-ray energy bands \citep{porter_high-energy_2017,galprop_v57,vladimirov_galprop_2011}.
To evaluate \acf{gde}, we make use of \galprop\ (version 54 of WebRun), which is designed to calculate astrophysics of \acp{cr} (i.e., propagation and energy loss) and photon emissions in the radio to gamma-ray energy bands \citep{porter_high-energy_2017,galprop_v57,vladimirov_galprop_2011}.
%\textcolor{blue}{\sout{A web-browser-based usage, WebRun, is also available \citep{vladimirov_galprop_2011}.} YI: We can remove this sentence.}
%%% Models in the literature.
%\textcolor{blue}{YI: We do not need a break here.}
In this paper, we consider three models of \ac{gde} in the literature:
one model from \cite{ackermann_fermi_2012} and
two models from \cite{orlando_imprints_2018}
(see Appendix \ref{sec:gde_appendix} and \tabref{tab:models} for details\footnote{Galdef files of the models are available in \url{https://tsuji703.github.io/MeV-All-Sky}.}).
%%% Ackermann+ 2012
%galdef_54_SNR_z4kpc_R20kpc_Ts150K_EBV5mag
These models are constructed to be reconciled with the gamma-ray observations by \lat\ and the observed \ac{cr} spectra by several \ac{cr} experiments.
%%% Orlando
Using the results of the latest \ac{cr} measurements with \voyager,
\cite{orlando_imprints_2018} modified the \ac{gde} models in the literature, especially the injection parameters of electrons and propagation parameters.
%%%
In this paper, a baseline model of
$^{\rm S}S ^{\rm Z}4 ^{\rm R}20 ^{\rm T}150 ^{\rm C}5$\footnote{This model assumes that the source distribution of \acp{cr} is SNRs, the Galactic disk is characterized by the height of $z=$4 kpc and the galactocentric radius of $R=$20 kpc, and $T_s$=150 K and $E(B-V) = 5 $ mag cut is adopted for determining the gas-to-dust ratio (see \citealt{ackermann_fermi_2012} for details).}
is selected as a representative of the models in \cite{ackermann_fermi_2012} and referred to as Model~1.
From \cite{orlando_imprints_2018}, we adopt the DRE (i.e., diffusion and re-acceleration) and DRELowV (modified DRE) models, hereafter referred to as Model 2 and Model~3, respectively.

\begin{figure}[ht!]
\plotone{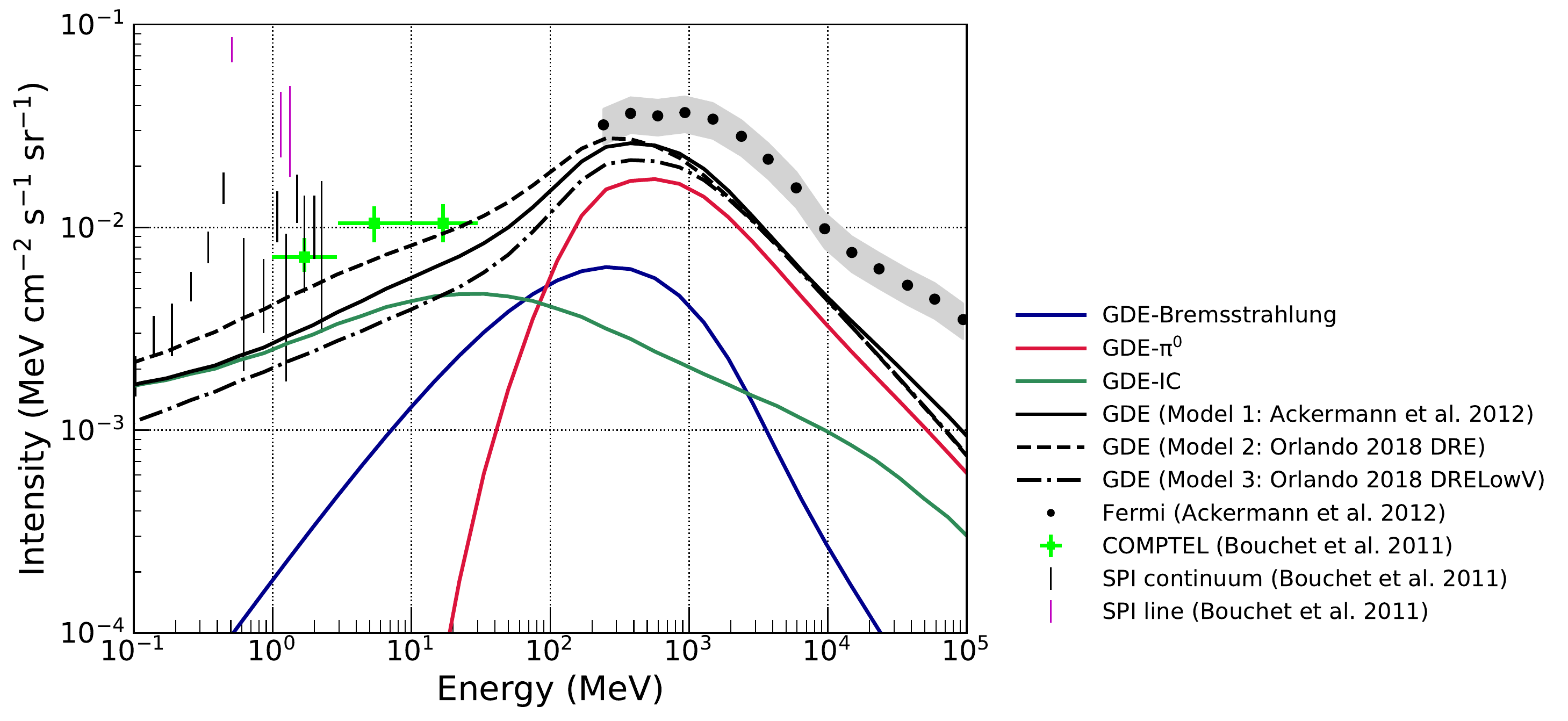}
%\plotone{figures/SED_detailed_mix_1h_noSource_v1.1.pdf}
%\plotone{figures/SED_detailed_mix_1h.pdf}
%\plotone{figures/SED_GDE_compared.png}
%\plotone{figures/GDE_comparison_v0.pdf}
\caption{
The spectra of the inner Galactic diffuse emission taken by \comptel\ in light green, shown with the continuum emission and gamma-ray lines by \spi\ in black and magenta, respectively \citep{bouchet_diffuse_2011}.
The components of \ac{gde} of Model~1 \citep{ackermann_fermi_2012} are shown in solid lines: total in black, Bremsstrahlung in blue, \pizero -decay in red, and \ac{ic} in green.
%The components of the Galactic diffuse emission (Model~1; \cite{ackermann_fermi_2012}) are shown in dashed lines; total in dashed black, Bremsstrahlung in dashed blue, \pizero -decay in dashed red, and \ac{ic} in dashed green.
%The source component is shown in red.
%The \ac{cgb} is shown in solid green line (\comptel; \eqref{eq:CGB}) and in magenta shadow.
The flux points observed by \lat, including the diffuse emission and the gamma-ray sources, are indicated by black circles with the grey shadow being the error, although it was obtained from the region with $|\ell| \leq 80\degr$ and $|b| \leq 8\degr$ \citep{ackermann_fermi_2012}.
%\check{Remove the all-sky image}
%\check{Add comparison of GDE models}
The total \ac{gde} of Model~2 and Model~3 \citep{orlando_imprints_2018} are illustrated with black dashed and dash-dotted lines, respectively.
%\comment{legend: change SPI inner to \integral-SPI}
\label{fig:fig1}
}
\end{figure}

%%% Example of DGE SED
%\figref{fig:fig1} shows \ac{gde} spectra of Model 1--3.
The \ac{gde} spectra of Model~1, shown in \figref{fig:fig1}, consist of three components of radiation: \ac{ic} scattering, Bremsstrahlung, and pion-decay radiation.
In the energy channel of \comptel, \ac{ic} is dominant, while Bremsstrahlung is subdominant because of ionization loss of electrons at the lower energy, and the hadronic component is less effective due to the pion bump.
The \ac{ic} scattering in the MeV gamma-ray range is attributed to sub-GeV electrons that up-scatter seed photon fields of optical, infrared, and \ac{cmb}.

%%% Compare GDE Model 1-3
\figref{fig:fig1} also compares the aforementioned GDE models, Models 1--3.
Since GDE below $\sim$100 MeV is dominated by the \ac{ic} component, the difference in Models 1--3 arises from \ac{cr} electrons in 0.1–1 GeV (\figref{fig:cr} in Appendix \ref{sec:gde_appendix}).
Models 2 and 3 are respectively the highest and lowest with the difference of a factor of few, and Model 1 is in the middle of them.
Model~2 almost can reach to the \comptel\ emission, while it is slightly lower than the flux at the lower energy bins. This trend is the same for the diffuse emission measured in the different sizes in \figref{fig:size} (Appendix \ref{sec:region_appendix}).
Although there is such uncertainty on the \ac{gde} models, $\gtrsim$30\% of the \comptel\ excess is contributed by GDE.

%\revise{one paragraph about the results of the region dependence? or in Appendix?}

\section{Results and Discussion}
\label{sec:discussion}
%%%%%%%%%%%%%%%%%%%%%%%%%
%%%%%%%%%%%%%%%%%%%%%%%%%

\figref{fig:model} compares the \comptel\ data points and our models with the three different models of \ac{gde}.
%We show \ac{gde} (Model~1 in \figref{fig:model} (a), Model~2 in (b), and  Model~3 in (c)), the spectrum of MeV gamma-ray sources, and \ac{cgb}, and combined spectra of these components.
%We show \ac{gde} Model~1--3, the spectrum of MeV gamma-ray sources, \ac{cgb}, and combined spectra of these three components.
We show the spectrum of MeV gamma-ray sources, \ac{gde} Models~1--3, and combined spectra of these two components.
It should be noted that this direct comparison did not take into account energy dispersion, which may have a significant effect on the result, as described in the data analysis of \spi\ \citep{strong_gamma-ray_2005}.
The issue might be large in the \comptel\ data due to its property of functioning as a Compton telescope.
Applying the energy response function, however, is beyond the scope of this study, and it is currently impossible because the \comptel\ response is not available.
Although a small ($\sim 1\sigma$) difference between the data points of \comptel\ and the models might not be counted as an excess unless the energy dispersion is properly considered,
in the following we report the results of the direct comparison between the data and the models.

%\comment{old ver.} Note that this direct comparison did not take into account energy dispersion, which may affect the result. However, this effect would be small because \comptel\ energy resolution of 5--10\% \citep{Comptel} is smaller than the energy bins of the \comptel\ emission, and applying the energy response function is beyond the scope of this study. Thus we compare the data points of \comptel\ and the models directly.

%Comments on each result.
We find that the combination of \ac{gde} and the sources can roughly reproduce the \comptel\ excess:
the entire spectrum can be sufficiently explained with Model~2 (\figref{fig:model} upper right), and
the lowest and highest energy bins of the \comptel\ data can be reproduced with Model~1 (\figref{fig:model} upper left).
In the case of Model~3 (\figref{fig:model} lower left), which is the GDE model with the smallest flux among the models considered in this paper, the combined spectrum of \ac{gde} and the sources is slightly lower than the \comptel\ excess.
%\sout{In all the cases, adding \ac{cgb} enables to reproduce the \comptel\ emission.}?
%\comment{or In the cases of Model~1 and Model~3, adding \ac{cgb} enables reproducing the \comptel\ emission (\figref{fig:cgb}).??}

%%% Region dependence
The inner Galactic diffuse emission is roughly $10^{-2}$ \Mflux\ for all the previous studies,
however the extracted region is different;
$|\ell| \leq 30\degr$ and $|b| \leq 15\degr$ in \cite{bouchet_diffuse_2011}, $|\ell| \leq 60\degr$ and $|b| \leq 10\degr$ in \cite{strong_diffuse_2004}, $|\ell| \leq 30\degr$ and $|b| \leq 5\degr$ in \cite{strong_comptel_1999}, $|\ell| \leq 45.7\degr$ and $|b| \leq 45.7\degr$ in \cite{siegert_diffuse_2022}, as illustrated in \figref{fig:size}.
This region difference is crucial for calculating the model.
%The results with the different regions are presented in Appendix \ref{sec:region_appendix}.
The trend in \figref{fig:model} is similar if we adopt for the region with  $|\ell| \leq 60\degr$ and $|b| \leq 10\degr$ or  $|\ell| \leq 30\degr$ and $|b| \leq 5\degr$ (see Appendix \ref{sec:region_appendix} for the results with the different regions).
In the region with $|\ell| \leq 45.7\degr$ and $|b| \leq 45.7\degr$, the model fails to reproduce the emission since it is extended to the high-latitude region, where \ac{gde} becomes faint and the number of sources decreases.

%%% Source spectrum
The spectral shape would provide us with a new constraint.
The power-law spectral index is $s \sim -1.9$ ($dN/dE \propto E^s$) for the \comptel\ excess from the inner Galaxy, while it is harder with  $s=-1.39$ for the latest measurement of the \spi\ emission in the region extended to the higher latitude \citep{siegert_diffuse_2022}.
\figref{fig:model} shows that the spectrum of the accumulated sources is almost flat in the SED with a spectral index of $s \sim -2$.
Since the \ac{gde} models have $s \sim -1.5$, the MeV gamma-ray sources should play an important role in reproducing the observed spectrum by \comptel\ with $s \sim -1.9$.
%Since the source flux is negligible compared to \ac{gde},
%More precise measurements of the source spectrum will be useful to determine the \ac{gde} spectrum, especially the index of the primary electrons responsible for the \ac{ic} radiation.
More precise measurements of the spectrum of each source will enable constraining the accumulative source spectrum, which in turn will be useful to determine the \ac{gde} spectrum, especially the index of the primary electrons responsible for the \ac{ic} radiation.

%\section{Discussion \comment{merge with 3. Results?} }
%%%%%%%%%%%%%%%%%%%%%%%%%
%%%%%%%%%%%%%%%%%%%%%%%%%

%%% Uncertainty of GDE
Since the dominant component in the energy range of the \comptel\ excess is \ac{gde}, the uncertainty of \ac{gde} prevents us from
reaching a robust conclusion.
The uncertainty of \ac{gde} arises from (1) the amount of \ac{cr} electrons and (2) the photon fields being up-scattered by the electrons. % through the \ac{ic} scattering.
%%% CR electron
The CR electrons in 100--1000 MeV, which produce 1--30 MeV photons via \ac{ic} scattering, are different by a factor of $\sim$4, depending on the models.
To distinguish these models is important in the perspective of \ac{cr} feedback on galaxy evolution:
\acp{cr} can produce a non-thermal pressure gradient and enhance the degree of ionization in molecular clouds, significantly affecting a star-forming activity
(e.g., \citealt{jubelgas_cosmic_2008,hopkins_first_2021,owen_observational_2021}).
%\textcolor{blue}{YI: \citep{Owen2021ApJ...913...52O}}\check{Any other preferred reference}.
%%% photon fields
%\comment{comment on the uncertainty of photon fields.}
The density of the photon fields across the Galaxy is not well constrained, which also makes the \ac{gde} model somewhat uncertain in the MeV gamma-ray band.
If we assume that the interstellar radiation field is locally enhanced, such as in the Galactic bulge or a large Galactic CR halo, \ac{gde} in the inner Galactic region is increased enough to reach the flux level of the \comptel\ excess \citep{bouchet_diffuse_2011}.

%%% inhomogeneity
%\comment{Cut out this paragraph?} \textcolor{blue}{YI: I do not think we need this paragraph considering the word limits. Just comment out.}
%In this paper, we consider the three models in the literature to account for the direct cosmic-ray data by \voyager\ and \ams\ and the all-sky GeV gamma-ray observation by \lat. We should be cautious about the difference between these two-type observations: while the cosmic-ray data are obtained in the vicinity of the Earth, the gamma-ray observation contains the information of the entire Galaxy. If the cosmic rays are not uniform through the Galaxy, the locally measured cosmic-ray spectra by \voyager\ and \ams\ do not necessarily have to be reproduced by the model, and the other model parameters may be allowed. However, the lack of the cosmic-ray observation outside our neighbor makes it difficult to evaluate the inhomogeneity. Furthermore, the current models are already fine-tuned, and thus changing the parameters might easily cause a discrepancy with the radio synchrotron and GeV gamma-ray maps.

%%% COMPTEL and Fermi-LAT GC excesses
%\comment{Comments on \lat\ GC excess?} \textcolor{blue}{YI: But, no space I think. Let's skip it. But, if the space allows, it may be interesting to discuss whether we can test the MSP scenarios for the GC excess in the MeV band.}

\subsection{Cosmic gamma-ray background}
\label{sec:cgb}
%%%%%%%%%%%%%%%%%%%%%%%%%
%%%%%%%%%%%%%%%%%%%%%%%%%
%\comment{move this subsection to Discussion section?}

%%% COMPTEL excess includes CGB or not?
%In this section,
%In addition to the sources and \ac{gde} (\secref{sec:sources} and \secref{sec:gde}) as primary components that must be accounted for the \comptel\ excess, we discuss how much it would be affected by uncertainty of subtraction of \ac{cgb} as a secondary component.
Here, we discuss the uncertainty on the subtraction of \ac{cgb}.
When calculating the \comptel\ emission, there is an isotropic term, $I_B$, in Equation (1) in \cite{strong_diffuse_1994,strong_diffuse_1996}, which likely corresponds to \ac{cgb}.
Later, \cite{strong_comptel_1999,strong_diffuse_2004,bouchet_diffuse_2011} presented the \comptel\ diffuse emission with \ac{cgb} being removed, as the base level (i.e., the zero-flux level) was set to the high-latitude sky.
We need to be cautious, however, of the treatment of the isotropic term:
First, since the isotropic term is a term with only the normalization being free, the spectral shape of \ac{cgb} was not taken into account.
Second, the uncertainty of the background subtraction might be included in the isotropic term, although the uncertainty of the overall fit was dominated by systematic errors, estimated to be of order of $\sim$25\%  \citep{strong_diffuse_1994}.
Therefore, there might be a possibility that the isotropic term could not completely represent \ac{cgb}.

\ac{cgb} in the MeV gamma-ray band derived from \comptel\ \citep{weidenspointner_cosmic_2000,kappadath_total_1997,
kappadath_measurement_1998} is reproduced by a broken power-law model:
\begin{equation}
    I(E) = 2.2 \times 10^{-4} \left(\frac{E}{3~\mathrm{MeV}} \right)^{-\Gamma} ~  (\mathrm{MeV ~ cm}^2 ~\mathrm{s ~ sr})^{-1} ,
    %% case
    %    \left\{
    %\begin{array}{ll}
    % 2.2 \times 10^{-4} \left(\frac{E}{3~\mathrm{MeV}} \right)^{-3.3}  (E \leq 3~\mathrm{keV}) \\
    %E^{-*}  (E> 3~\mathrm{keV}) \\
    %\end{array} \right.
    \label{eq:CGB}
\end{equation}
where the spectral slope $\Gamma$ is 3.3 for $E \leq 3$ MeV and 2 for $E > 3$ MeV.
This is roughly comparable with the observation of the Solar Maximum Mission (SMM) Gamma-Ray Spectrometer (GRS) \citep{watanabe_mev_2000}. % within a few factor.
%In this paper, we use the \comptel\ model (\eqref{eq:CGB}) for estimation of \ac{cgb}.
%\figref{fig:model} shows that \ac{cgb} is slightly larger than the source component and has $\gtrsim$20\% flux of the \comptel\ excess.
%
%
%%% uncertainty of CGB subtraction
A fraction of CGB (\eqref{eq:CGB}) to the \comptel\ excess is approximately 20\%, except for the lowest energy bin of which the fraction is 60\%.
Since this fraction at the higher energy bins is smaller than the systematic uncertainty of 25\%, the \comptel\ data points include the uncertainty of the subtraction of CGB.

%%% Adding CGB
\if0
\comment{cut this paragraph?}
%\comment{multiple CGB by 25\% because of the systematic uncertainty (is this correct?) or show **\% is enough to fill the gap.}
\revise{
Using Model~2, the \comptel\ emission can be reproduced by the GDE and sources, and additional components are not needed.
In the cases of Model~1 and Model~3, adding $\sim$45\% and $\sim$65\% of CGB (\eqref{eq:CGB}) enables to explain the \comptel\ excess, respectively (\figref{fig:cgb}).
% In the cases of Model~1 and Model~3, adding \ac{cgb} enables reproducing the \comptel\ emission (\figref{fig:cgb}).??}
}

\begin{figure}[ht!]
\plotone{figures/SED_1panel_Ackermann+2012_30.0x15.0_binTrue_CGBfactor0.46_addUNmatchTrue.pdf} \plotone{figures/SED_1panel_Orlando2018DRELowV_30.0x15.0_binTrue_CGBfactor0.65_addUNmatchTrue.pdf}
%\plottwo{figures/SED_1panel_Ackermann+2012_30.0x15.0_binFalse_CGBfactor0.46_addUNmatchTrue.pdf}{figures/SED_1panel_Orlando2018DRELow_30.0x15.0_binFalse_CGBfactor0.65_addUNmatchTrue.pdf}
\caption{
\revise{
Same as \figref{fig:model} of Model~1 and Model~1,
but CGB is added.
}
\label{fig:cgb} }
\end{figure}

\fi

\subsection{Prospects for future missions}
%%%%%%%%%%%%%%%%%%%%%%%%%
%%%%%%%%%%%%%%%%%%%%%%%%%

%%% Future mission.
%Towards an advanced understanding of the inner Galactic diffuse emission, observations with much better performance (i.e., greater angular resolution and larger effective area) are desired yet have not been achieved for more than two decades since \comptel.
In order to have an advanced understanding of the inner Galactic diffuse emission, observations with much better performance (i.e., greater angular resolution and larger effective area) are desired.
Such observations have not been achieved in the two decades since \comptel.
There are several ongoing or planned projects of the MeV gamma-ray observation,
such as COSI-SMEX \citep{COSI_SMEX} to be launched as a satellite in 2026,
e-ASTROGAM \citep{eASTROGAM2018}, AMEGO-X \citep{fleischhack_amego-x_2021}, GRAMS \citep{Aramaki2020}, SMILE-3 \citep{takada_first_2022}, GECCO \citep{moiseev_new_2021}, and a CubeSat for MeV observations (MeVCube) \citep{lucchetta_introducing_2022}.
With these future missions, we need to resolve the individual MeV gamma-ray sources first.
In the inner Galactic region ($|\ell| \leq 30\degr$ and $|b| \leq 15$\degr), our source model predicts that there are 5, 26, and 28 sources with the flux in 1--10 MeV being larger than
$10^{-10}$, $10^{-11}$, and $10^{-12}$ \flux, respectively.
%In the inner Galactic region ($|\ell| \leq 60\degr$ and $|b| \leq 10$\degr), our source model predicts that there are 8, 32, and 36 sources with the flux in 1–10 MeV being larger than $10^{-10}$, $10^{-11}$, and $10^{-12}$ \flux, respectively.
These sources can be detectable by the observatory whose sensitivity is improved by 1--2 orders of magnitude from \comptel.
After subtracting the source contribution and \ac{cgb}, we can constrain \ac{gde} with higher accuracy, %, combined with the cosmic-ray, GeV gamma-ray, and synchrotron radio data.
then we can clarify the presence of the \comptel\ excess.

%%% other scenarios
Although more investigation is necessary to reveal whether there indeed is an excess in the inner Galactic diffuse emission in the MeV gamma-ray energy band, we address the other scenarios besides \ac{gde}, the sources, and \ac{cgb}.
For example, as indicated by Model~3, we would need additional component(s) to reconcile with the \comptel\ emission.
Possible explanations are low-mass ($\lesssim$280 MeV) annihilating dark matter coupling to first-generation quarks \citep{boddy_indirect_2015,christy_indirect_2022} and/or
cascaded gamma rays accompanying cosmic neutrinos \citep{fang_tev_2022},
which would open up a new window for these studies.

\section{Conclusion} \label{sec:conclusion}
%%%%%%%%%%%%%%%%%%%%%%%%%
%%%%%%%%%%%%%%%%%%%%%%%%%
To clarify the origin of the \comptel\ excess, we elaborated on models consisting of MeV gamma-ray objects and \ac{gde}. %, and supplementally \ac{cgb}.
The crossmatched sources (both the hard X-ray and GeV gamma-ray emitters) have contributions of $\sim$10\% to the \comptel\ diffuse emission, and the contribution of the unmatched sources  (either of the hard X-ray or GeV gamma-ray emitters) is also at the same level.
%\ac{gde} is dominant and the most uncertain, while the sources and \ac{cgb} have contribution of $\sim$10\% and $\sim$20\% to the \comptel\ diffuse emission, respectively.
Although the most uncertain component of \ac{gde} prevents us from a robust conclusion,
we found that the combination of all the components can roughly reproduce the \comptel\ excess, except for the \ac{gde} model with the smallest flux.
With future missions, we would be able to discriminate between the \ac{gde} models, enabling us to determine the amount of low-energy \ac{cr} electrons and characterize their role in the galaxy evolution, and confirm the existence of the \comptel\ excess, opening up a new window for dark matter or neutrinos if it exists.

%%%%%%%%%%%%%%%%%%%%%%%%%
%%%%%%%%%%%%%%%%%%%%%%%%%

%%%%%%%%%%%%%%%%%%%%%%%%%
%%%%%%%%%%%%%%%%%%%%%%%%%
\begin{acknowledgments}
We thank the anonymous referee for their advice, which was very helpful to improve our manuscript.
We also thank Dmitry Khangulyan, Nagisa Hiroshima, Susumu Inoue, Andrew W. Strong, and the GRAMS collaboration for the fruitful discussion.
%We thank the GRAMS collaboration.
This work made use of data from the \textit{Swift} and \fermi\ observatories and the \galprop\ code.
N.T. acknowledges support from the Japan Society for the Promotion of Science (JSPS) KAKENHI grant No. 22K14064.
H.Y. is supported by JSPS KAKENHI grant No. 20K22355,
Y.I. is supported by JSPS KAKENHI grant %Nos. JP16K13813, JP18H05458, and JP19K14772,
Nos. 18H05458 and 19K14772, %and 22K18277,
and H.O. is supported by JSPS KAKENHI grant Nos. %19H05185 and 19H01906.
19H01906, 19H05185, and 22H00128. %, 22K18277.
This work was partially supported by JSPS KAKENHI grant No. 22K18277.

\end{acknowledgments}


%%%%%%%%%%%%%%%%%%%%%%%%%
%%%%%%%%%%%%%%%%%%%%%%%%%
\vspace{5mm}

\facilities{
\bat, \lat
}

\software{
\galprop\ \citep{porter_high-energy_2017}\footnote{\url{http://galprop.stanford.edu}}
}

%\software{astropy \citep{2013A&A...558A..33A,2018AJ....156..123A}, Cloudy \citep{2013RMxAA..49..137F},           Source Extractor \citep{1996A&AS..117..393B}          }


%   Appendix
%%%%%%%%%%%%%%%%%%%%%%%%%
%%%%%%%%%%%%%%%%%%%%%%%%%

\clearpage
%%%%%%%%%%%%%%%%%%%%%%%%%
%%%%%%%%%%%%%%%%%%%%%%%%%
\appendix

\if0
%%% Notes
%%%%%%%%%%%%%%%%%%%%%%%%%
%%%%%%%%%%%%%%%%%%%%%%%%%
\begin{itemize}

\item {\bf Notes}
\begin{itemize}
    \item Less than 3,500 words, not including Acknowledgement, Appendix, and supplementary materials.
    \item Less than 5 figures and tables.
    \item References should be less than 50.
\end{itemize}

\item {\bf Important references}
\begin{itemize}
\item \comptel\ excess: 
%\cite{strongDiffuseContinuumGamma1994, strongDiffuseGalacticHard1996}
\cite{strong_diffuse_1994,strong_diffuse_1996}

\item Fermi gamma-ray diffuse and modeling:  %\cite{ackermann_fermi_2012}
\cite{ackermann_fermi_2012}

\item CR propagation model to fit \ams\ and \voyager\ data and MeV diffuse emission: %\cite{orlando_imprints_2018}
\cite{orlando_imprints_2018}

\item \integral -SPI diffuse: %\cite{bouchet_diffuse_2011,siegert_diffuse_2022}.
\cite{bouchet_diffuse_2011,siegert_diffuse_2022}
\end{itemize}

\end{itemize}

\fi



%%%%%%%%%%%%%%%%%%%%%%%%%%%%%%%%%%
%%%%%%%%%%%%%%%%%%%%%%%%%%%%%%%%%%
\section{Example SEDs of sources}
\label{sec:sources_appendix}

\begin{figure}[ht!]
\plottwo{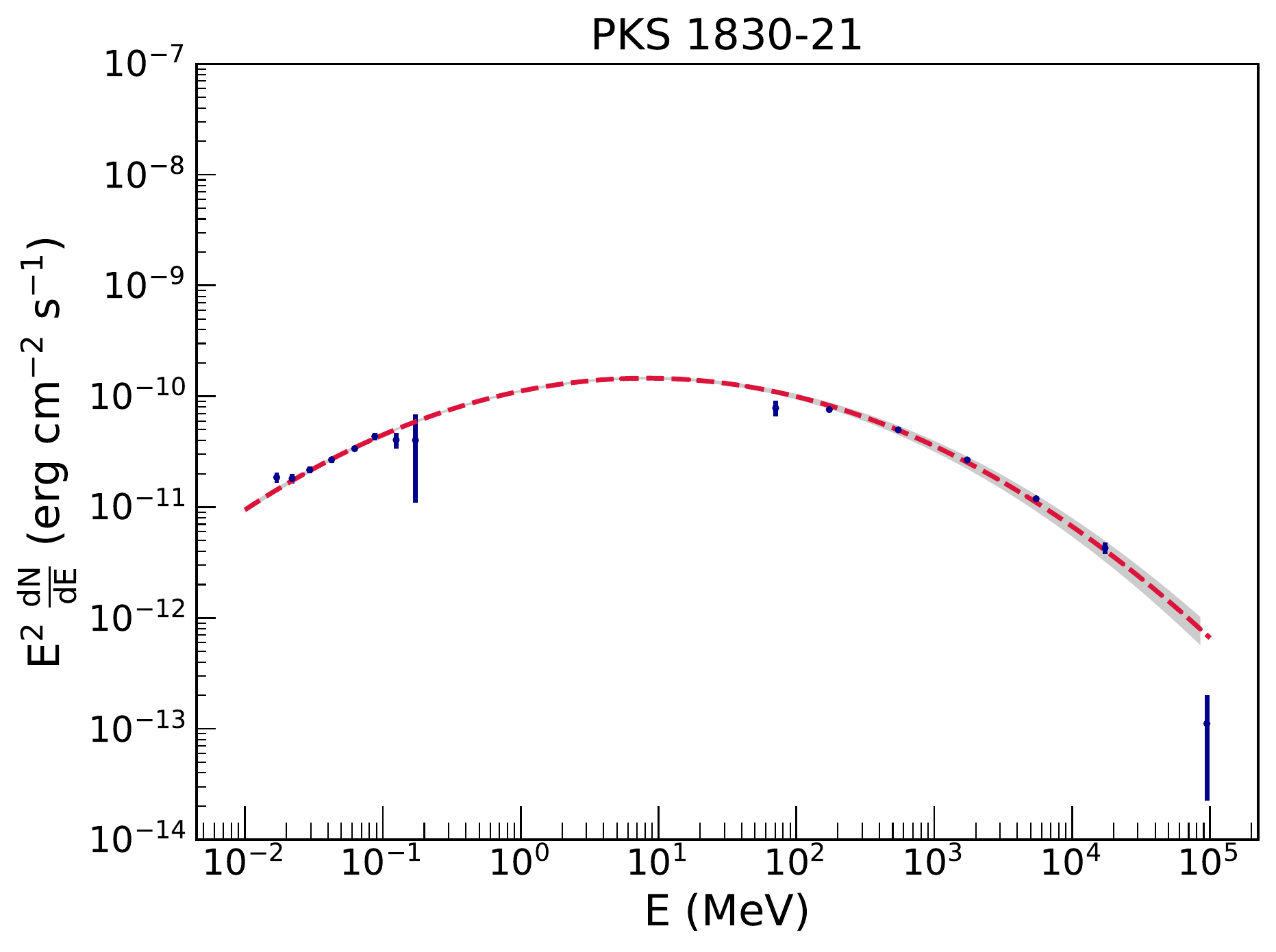}{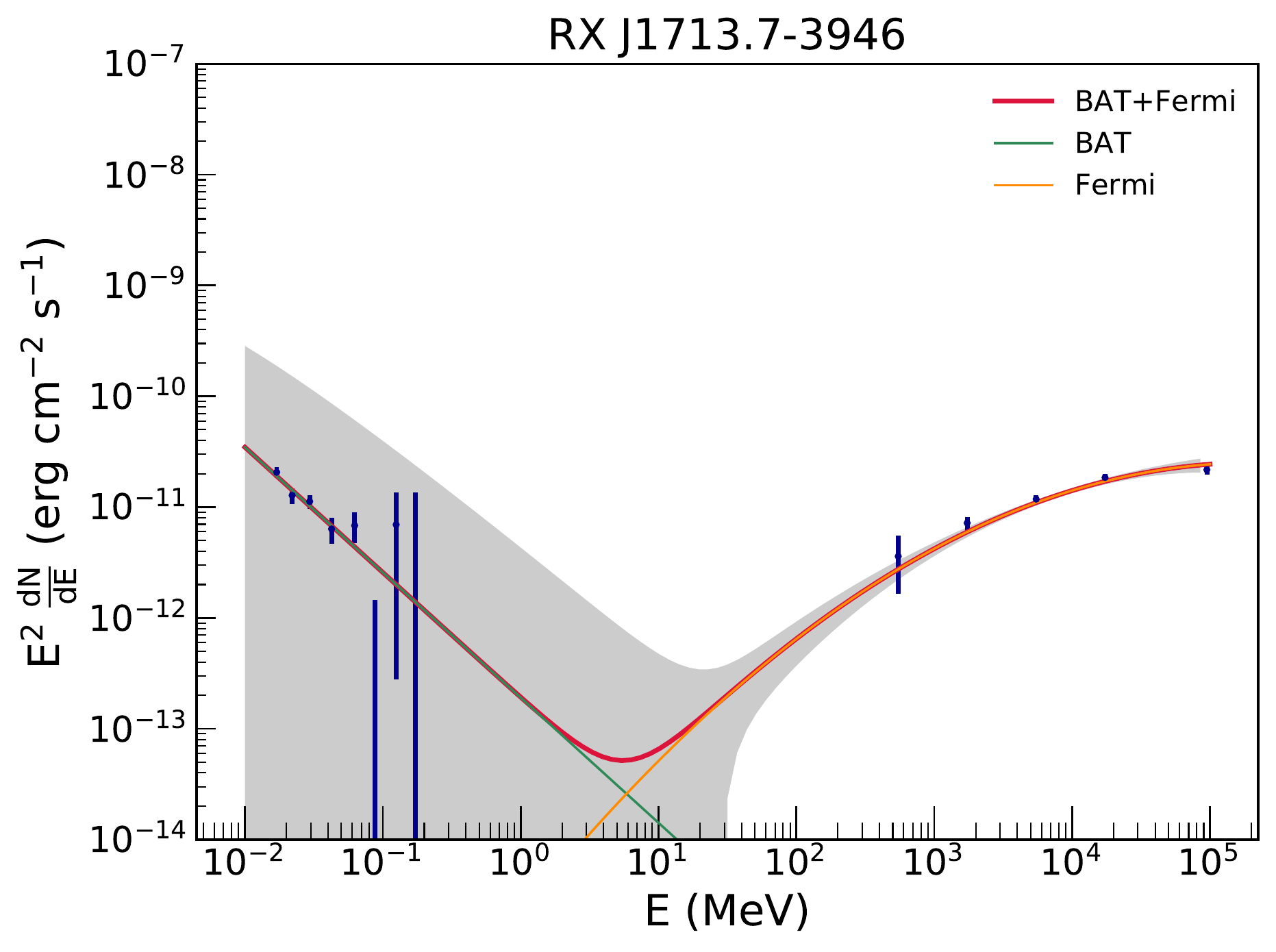} \\
%\plottwo{figures/fig_1RXS_J174036.3+521155.pdf}{figures/fig_Kes_73.pdf} \\%\plottwo{figures/fig_4C_+51.37.pdf}{figures/fig_Kes_73.pdf} \\
\plottwo{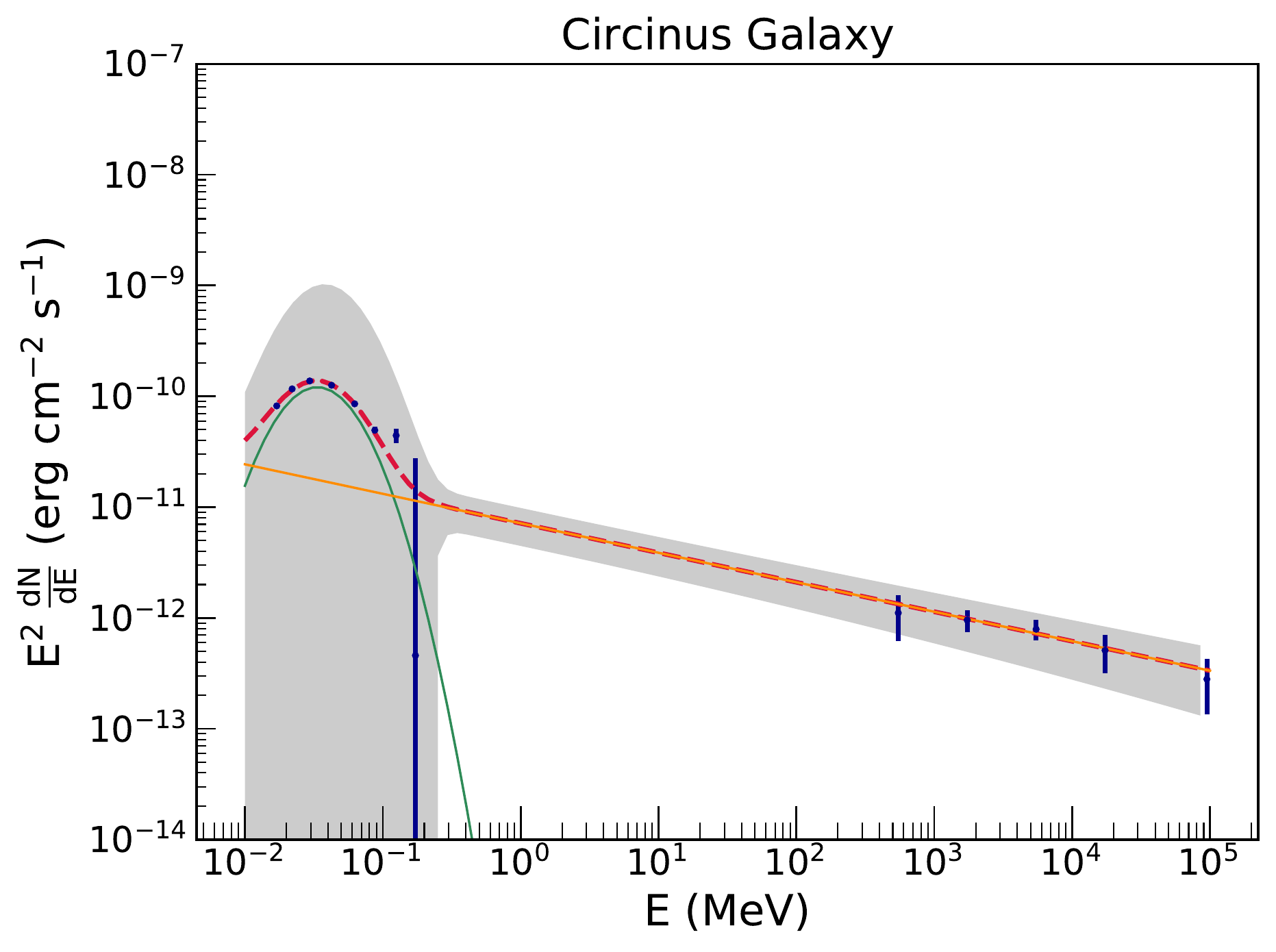}{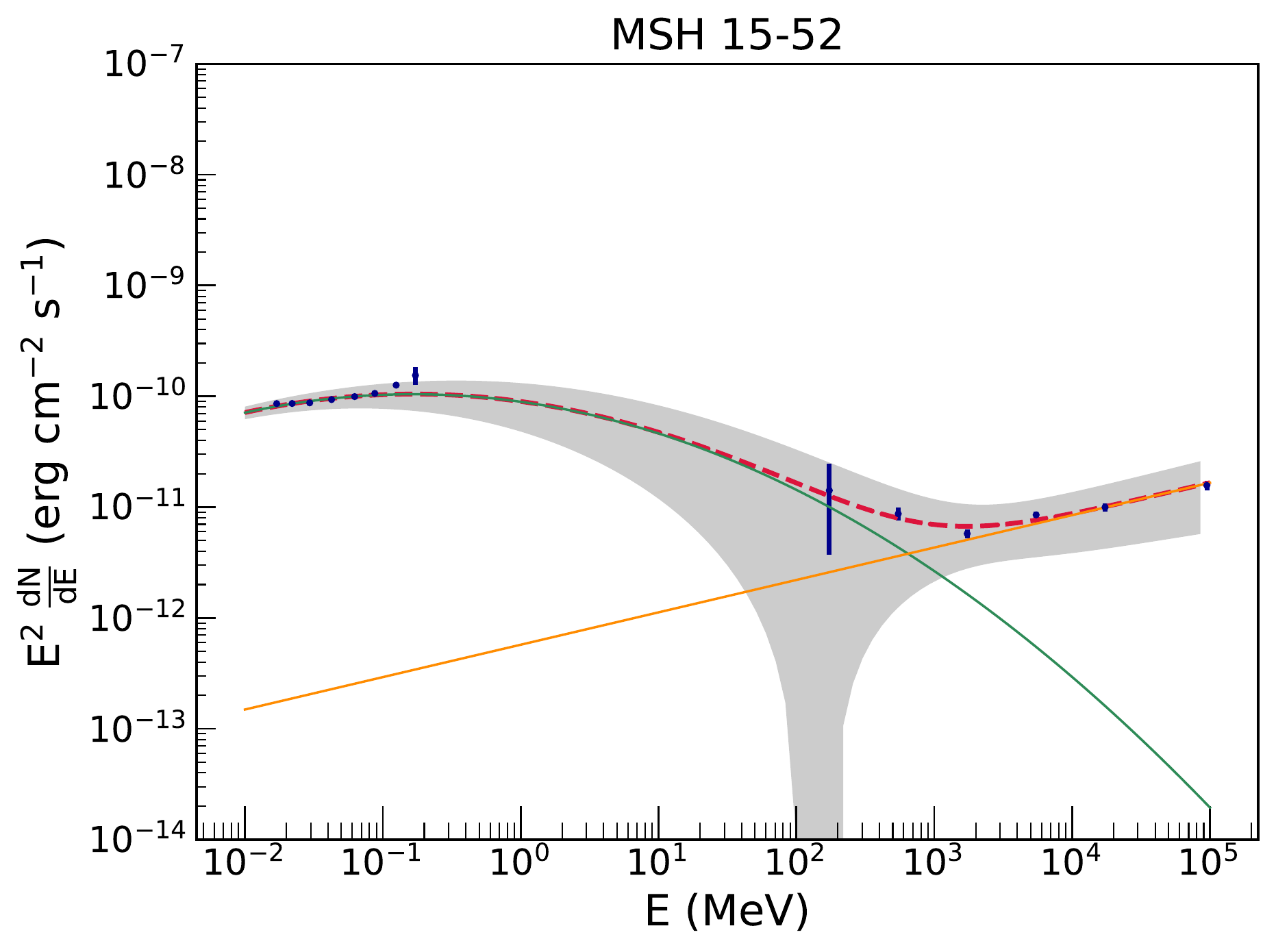}
\caption{
Example \acp{sed} of PKS 1830$-$21, \rxj, Circinus Galaxy, and MSH 15$-$52.
The flux points obtained by \bat\ and \lat\ are shown in blue.
The best-fit models (log parabola for PKS 1830$-$21, two-component model for \rxj, and source-dependent model for Circinus Galaxy and MSH 15$-$52) are illustrated with red lines. For the two-component model, each component is also shown in green and orange.
\label{fig:sed}
}
\end{figure}

\figref{fig:sed} shows the example \acp{sed} of the MeV gamma-ray sources PKS 1830$-$21, \rxj, Circinus Galaxy, and MSH 15$-$52.
%1RXS J174036.3$+$521155 ,  Kes 73,
The log-parabola model is applied to PKS 1830$-$21,
the two-component model is applied to \rxj\ (see \secref{sec:sources} for details),
and we adopt the source-dependent model for Circinus Galaxy and MSH 15$-$52.
We provide \acp{sed} of all of the MeV gamma-ray sources of \cite{tsuji_cross-match_2021} in
\url{https://tsuji703.github.io/MeV-All-Sky}.
%\url{https://github.com/tsuji703/MeV-All-Sky}.
%\comment{put this information in Zenodo?}

In the case of the extraction region with $|\ell| \leq 60\degr$ and $|b| \leq 10\degr$, Circinus galaxy and MSH 15$-$52 cannot be well-fit by neither log-parabola nor two-component models.
%The remaining two sources, Circinus galaxy and MSH 15$-$52, 
Therefore, they are fitted with a source-dependent model as follows (\figref{fig:sed}). %The hard X-ray emission of Circinus galaxy (a Seyfert galaxy) is reproduced by \textcolor{blue}{\sout{ MyTorus (Gaussian-like) model\footnote{https://www.mytorus.com/}} accretion disk emission}, while its gamma-ray emission is unknown \citep{hayashida_discovery_2013}. 
The hard X-ray emission of Circinus galaxy (a Seyfert galaxy) is reproduced by accretion disk emission, while its gamma-ray emission is unknown \citep{hayashida_discovery_2013}. 
We apply a combination of a Gaussian model in the hard X-ray band and a power-law model from 4FGL. 
The emission in the hard X-ray to GeV gamma-ray bands from MSH 15$-$52 could be given by a superposition of emission from the central pulsar and its nebula, for which we adopt log-parabola and power-law models, respectively \citep{abdo_detection_2010}.

%\section{Table of sources}
%\check{Present a table to list all the sources in the inner Galactic region}

%%%%%%%%%%%%%%%%%%%%%%%%%%%%%%%%%%
%%%%%%%%%%%%%%%%%%%%%%%%%%%%%%%%%%
\section{Comparison of GDE models}
\label{sec:gde_appendix}

%\comment{move the main text?}

\begin{figure}[ht!]
%\plotone{figures/GDE_comparison_CR.png}
\plottwo{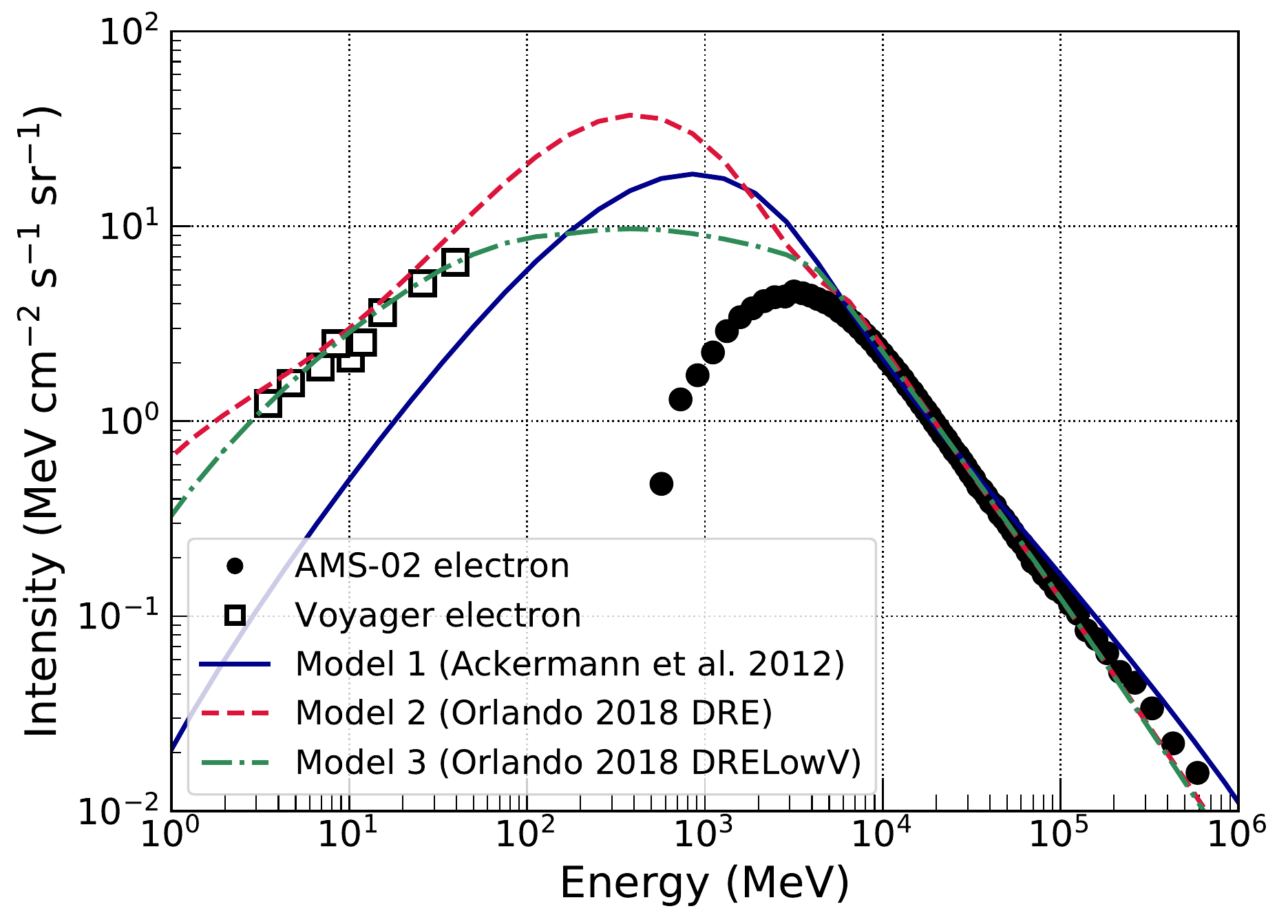}{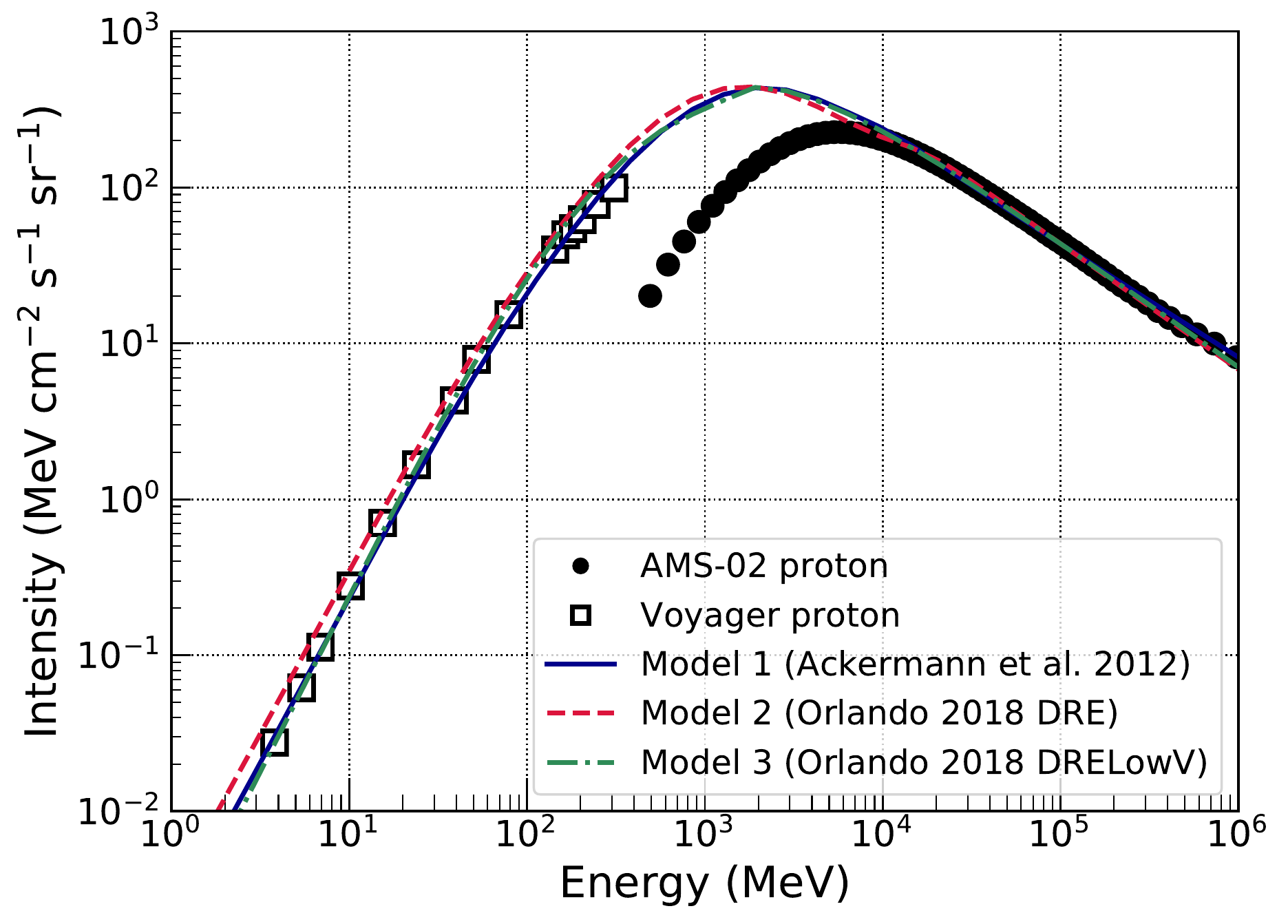}
%\plotone{figures/GDE_comparison_v0.pdf}
\caption{
%Comparison of \ac{cr} spectra of electrons (left) and protons (right), obtained around the Sun ($R$=8 kpc from the Galactic center) for the different GDE models (Model~1 \citep{ackermann_fermi_2012} in blue; Model~2 in red; and Model~3 in green \citep{orlando_imprints_2018}).
Comparison of \ac{cr} spectra of electrons (left) and protons (right) around the Sun ($R$=8 kpc from the Galactic center) for the different GDE models: Model~1 taken from \cite{ackermann_fermi_2012} in blue; Model~2 from \cite{orlando_imprints_2018} in red; and Model~3 from \cite{orlando_imprints_2018} in green.
The black filled circles and open squares indicate the \ac{cr} data of AMS-02 \citep{AMS2014_ep,AMS2015_proton} and \voyager\ \citep{cummings_galactic_2016}, respectively.
%\comment{The bottom panel is not necessary because it is the same as Figure~1.}
\label{fig:cr}
}
\end{figure}

This section gives brief descriptions of each GDE model.
The characteristic parameters of the \ac{gde} models of Models 1--3 are summarized in \tabref{tab:models}.
\figref{fig:cr} shows a comparison of \ac{cr} spectra around the Solar System, calculated by the \ac{gde} models in \tabref{tab:models}.
The observed \ac{cr} spectra by \voyager\ and \ams\ are also illustrated. Note that the spectra of \ams\ are affected by solar modulation below a few GeV, and this solar modulation is not accounted for the shown \ac{gde} models.

%%% Ackermann+ 2012
%galdef_54_SNR_z4kpc_R20kpc_Ts150K_EBV5mag
\cite{ackermann_fermi_2012} searched for parameters of \ac{gde} which are reconciled with the gamma-ray observations by \lat\ and the observed \ac{cr} spectra by several \ac{cr} experiments (ACE, AMS, JACEE, HEAO-3, BESS, CREAM, and ISOMAX).
% Ackermann+ 12, Section 3.2: BESS, ACE, AMS, JACEE, HEAO-3. Figure 31: CREAM, ACE, ISOMAX for B/C and 10Be/9Be ratio.
Their models successfully reproduce both the spectral and spatial distributions observed by \lat\ in the 0.1--100 GeV gamma-ray band. 
In this paper, a baseline model of $^{\rm S}S ^{\rm Z}4 ^{\rm R}20 ^{\rm T}150 ^{\rm C}5$
%\footnote{This model assumes that the source distribution of \acp{cr} is SNRs, the Galactic disk is characterized by the height of $z=$4 kpc and the galactocentric radius of $R=$20 kpc, and $T_s$=150 K and $E(B-V) = 5 $ mag cut is adopted for determining gas-to-dust ratio (see \cite{ackermann_fermi_2012}) for details).} 
is selected as a representative of the models in \cite{ackermann_fermi_2012} and referred to as Model~1. %\textcolor{blue}{YI changed orders of sentences in this and the next paragraph.}

%%% Orlando 2018
The models, however, are constructed to be in agreement with the local \ac{cr} spectra, causing a discrepancy with the sub-GeV \ac{cr} spectra outside the Solar System taken by \voyager.
Particularly, the spectral model of \ac{cr} electrons turned out to be lower than the \voyager\ data by a factor of 4--10 (\figref{fig:cr}).
Taking into consideration the latest measurements of \acp{cr} with \voyager,
\cite{orlando_imprints_2018} modified the \ac{gde} models in the literature, especially the injection parameters of electrons and propagation parameters.
The new models in \cite{orlando_imprints_2018} are consistent with both the \ac{cr} and gamma-ray observations.
Among them, we adopt models of DRE (i.e., diffusion and re-acceleration) and DRELowV (modified DRE), respectively corresponding to the models with the largest and lowest %\comment{compared with what?} 
amount of \ac{cr} electrons in the energy range of 0.1–1 GeV \citep{orlando_imprints_2018}. 
%Hereafter, t
The DRE and DRELowV models are respectively referred to as Model 2 and Model~3.
Note that the spectral model of \ac{cr} proton in Models~2--3 is roughly comparable with Model 1, while that of electron is higher below GeV. % (see \secref{sec:gde_appendix}).

\begin{deluxetable*}{ cccc}
%\tablenum{1}
\tablecaption{
\ac{gde} models.
%\comment{Cut or move it to Appendix}
%%% GALPROP. 
%				run#1h = Ackermann12, 
%				run#1j = Orlando18 DRE, 
%				run#1i= Orlando18 DREVlow
\label{tab:models}
}
\tablewidth{0pt}
\tablehead{
\nocolhead{} & \colhead{Model 1} & \colhead{Model 2} & \colhead{Model 3} 
%\multicolumn2c{Distance} & \colhead{} & \colhead{V} \\
%\colhead{Number} & \colhead{Number} & \nocolhead{Name} & \colhead{Type} &
%\multicolumn2c{(kpc)} & \colhead{Constellation} & \colhead{(mag)}
}
%\decimalcolnumbers
\startdata
    Ref. & \cite{ackermann_fermi_2012} & \cite{orlando_imprints_2018} DRE & \cite{orlando_imprints_2018} DRELowV  \\  
    \hline
%    \multicolumn{4}{c}{Propagation} \\ \cline{2-4}
    Propagation & & & \\  \cline{1-1}
    $v_{\rm Alf}$ & 33 & 38 & 20  \\
    $D0_{\rm xx}$ (10$^{28} ~ {\rm cm}^2~{\rm s}^{-1}$) & 5.2 & 15 & 15 \\
    $D_{\rm br}$ (GV) & 4.0 & 40 & 40 \\
    $\gamma_{1}$ & 0.33 & 0.33 & 0.33 \\
    $\gamma_{2}$ & 0.33 & 0.33 & 0.33 \\
    \hline
%    \multicolumn{4}{c}{Proton / Nucleus} \\ \cline{2-4}
    Injection (Proton \& Nucleus) & & &  \\ \cline{1-1}
    $E_{\rm br}$ (GV) & 11 & 18 & 2.7  \\
    $s_1$ & 1.9 & 1.9 & 1.4  \\
    $s_2$ & 2.4 & 2.5 & 2.5 \\
    \hline
%    \multicolumn{4}{c}{Electron} \\ \cline{2-4}
    Injection (Electron) & & & \\ \cline{1-1}
    $E_{\rm br, 0}$ (MV) & 2.2$\times 10^3$ & 320 & 170  \\
    $E_{\rm br, 1}$ (GV) & 2.2$\times 10^3$ & 6.3 & 4.5  \\
    $s_0$ & 1.6 & 2.9 & 2.2 \\
    $s_1$ & 2.4 & 0.80 & 1.7  \\
    $s_2$ & 4.0 & 2.7 & 2.7 \\
\enddata
\tablecomments{
%This table ``hides'' the third column in the \latex\ when compiled.
The diffusion coefficient is given by $D(\rho)= \beta D0_{\rm xx} ( \rho / D{_\mathrm{br}} )^ \gamma$, where $\beta=v/c$ and $\rho=cp/(Ze)$.
}
\end{deluxetable*}

%%%%%%%%%%%%%%%%%%%%%%%%%%%%%%%%%%
%%%%%%%%%%%%%%%%%%%%%%%%%%%%%%%%%%
\section{Dependence on region}
\label{sec:region_appendix}

\begin{figure}[ht!]
%\plotone{figures/SED_mix_grid_v1.1_Gal_60.0x10.0_binTrue_CGBfactor0_addUNmatchTrue.pdf}
\centering \includegraphics[height=0.5\hsize]{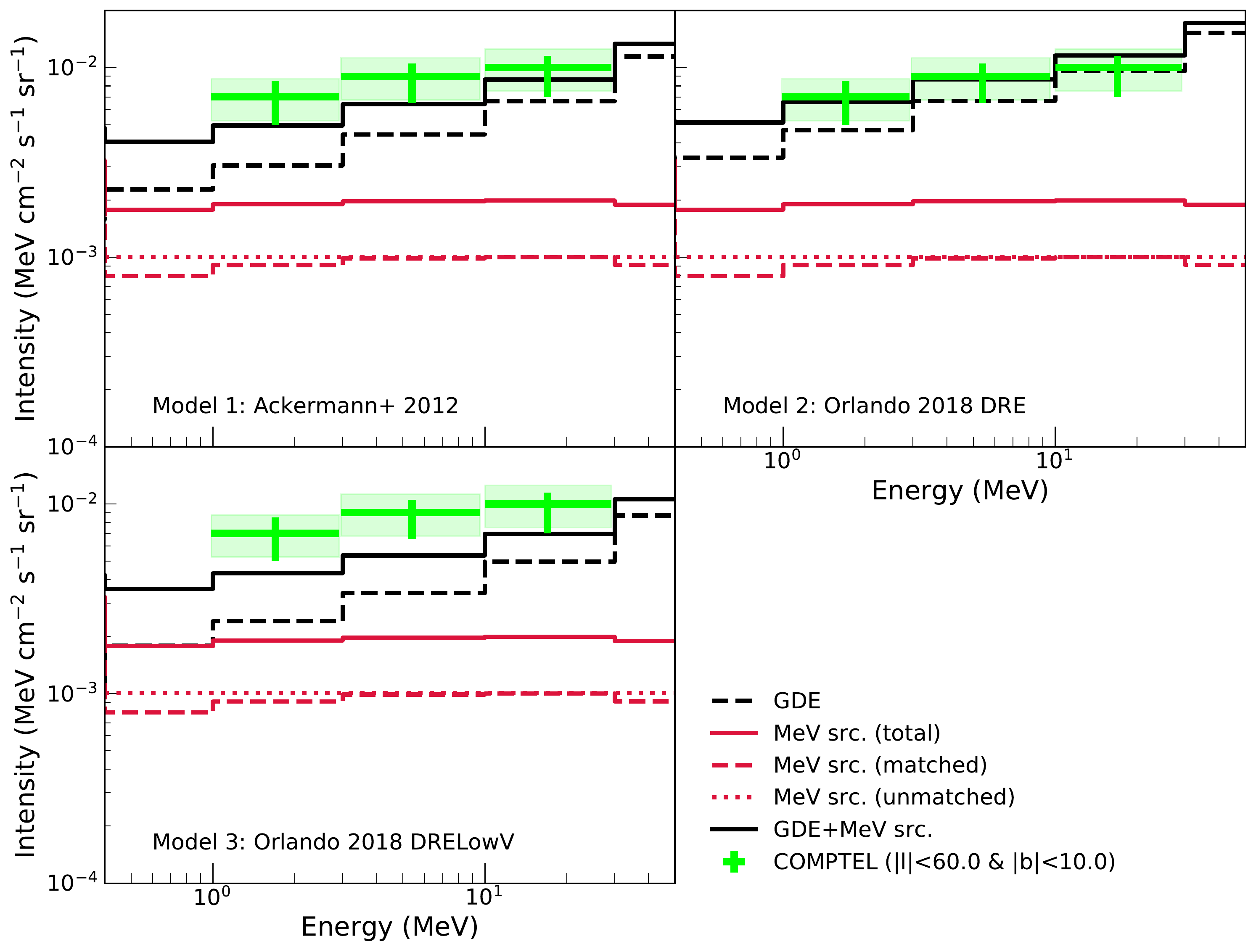}
\caption{
Same as \figref{fig:model}, but calculated for $|\ell| \leq 60\degr$ and $|b| \leq 10\degr$ \citep{strong_diffuse_1996}.
\label{fig:region1}
}
\end{figure}

\begin{figure}[ht!]
%\plotone{figures/SED_mix_grid_v1.1_Gal_30.0x5.0_binTrue_CGBfactor0_addUNmatchTrue.pdf}
\centering \includegraphics[height=0.5\hsize]{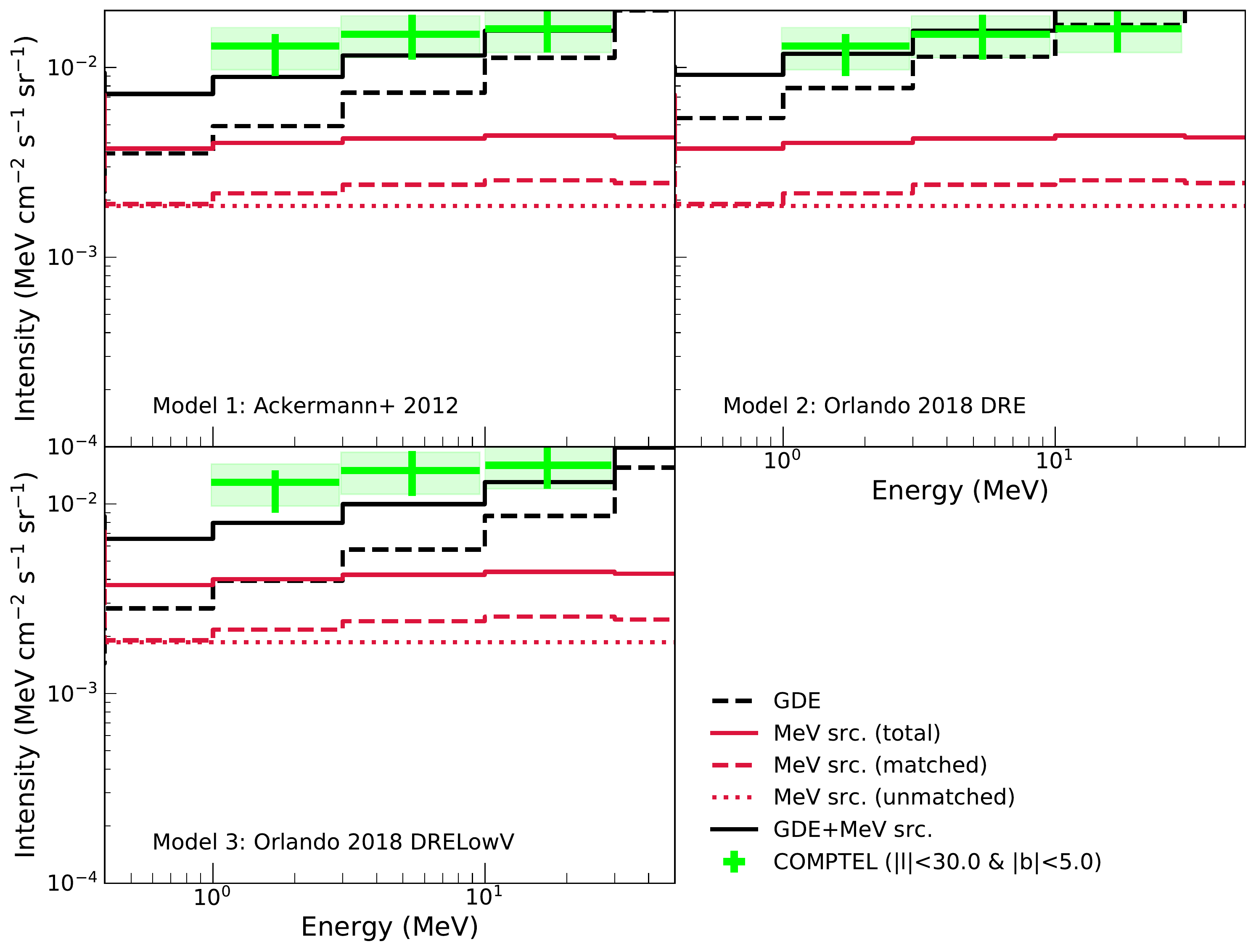}
\caption{
Same as \figref{fig:model}, but calculated for $|\ell| \leq 30\degr$ and $|b| \leq 5\degr$ \citep{strong_comptel_1999}.
\label{fig:region2}
}
\end{figure}

\begin{figure}[ht!]
%\plotone{figures/SED_mix_grid_v1.1_Gal_47.5x47.5_binTrue_CGBfactor0_addUNmatchTrue.pdf}
\centering \includegraphics[height=0.5\hsize]{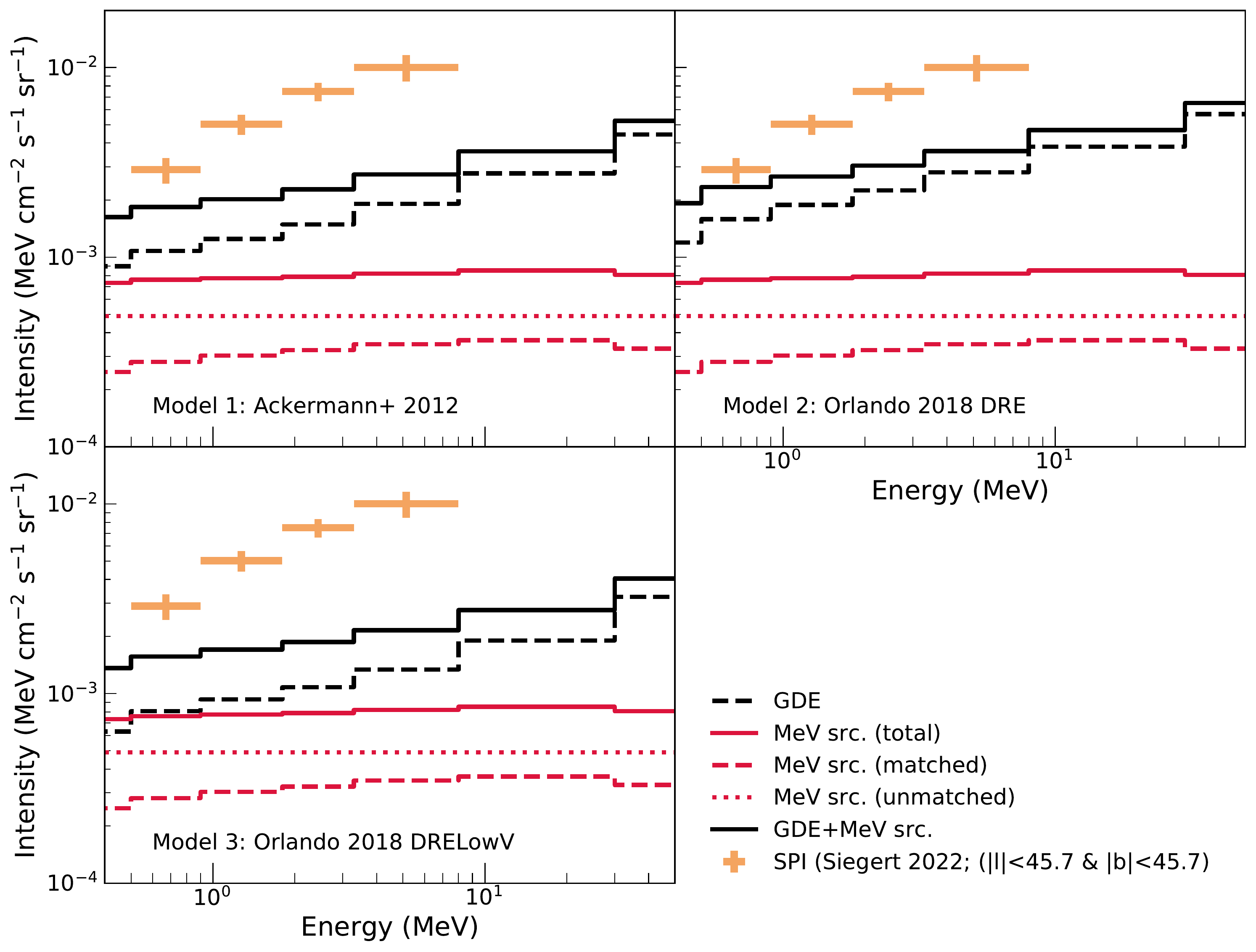}
\caption{
Same as \figref{fig:model}, but calculated for $|\ell| \leq45.7\degr$ and $|b| \leq45.7\degr$ \citep{siegert_diffuse_2022}.
\label{fig:region3}
}
\end{figure}

This section presents the results for the different regions, $|\ell| \leq60\degr$ and $|b| \leq10\degr$ \citep{strong_diffuse_1996},
$|\ell| \leq30\degr$ and $|b| \leq5\degr$ \citep{strong_comptel_1999},
and $|\ell| \leq45.7\degr$ and $|b| \leq45.7\degr$ \citep{siegert_diffuse_2022}, respectively shown in \figref{fig:region1}, \figref{fig:region2}, and \figref{fig:region3}.
%\comment{mention how much, in \%?, the MeV sources contribute to the emission.}
The results are summarized in \tabref{tab:region}.
The numbers of matched and unmatched sources are 40 and 1162 in the region with $|\ell| \leq60\degr$ and $|b| \leq10\degr$,
25 and 551 with $|\ell| \leq30\degr$ and $|b| \leq5\degr$,
and 46 and 1923 with $|\ell| \leq45.7\degr$ and $|b| \leq45.7\degr$.
In the cases of the former two regions, %$|\ell| \leq60\degr$ and $|b| \leq10\degr$ (\cite{strong_diffuse_1996} and \figref{fig:region1}) and $|\ell| \leq30\degr$ and $|b| \leq5\degr$ (\cite{strong_comptel_1999} and \figref{fig:region2}), 
the total (matched plus unmatched) source contribution to the \comptel\ emission is 20--30\%,
and the \comptel\ emission can be roughly reproduced by a combination of \ac{gde} and the sources, likewise the region with $|\ell| \leq30\degr$ and $|b| \leq15\degr$ (\figref{fig:model}).
In the case of $|\ell| \leq45.7\degr$ and $|b| \leq45.7\degr$ (\citealt{siegert_diffuse_2022} and \figref{fig:region3}), the total source contribution is 8--25\% with respect to the \spi\ emission,
and there is an apparent difference between the emission and the model.
In such a high-latitude region, \ac{gde} becomes faint and the number of sources decreases. Especially, there is no contribution of extended (i.e., Galactic) sources.

\begin{deluxetable*}{ cccccc }
\tablecaption{
Summary of the MeV gamma-ray inner Galactic diffuse emission and sources in different regions.     \label{tab:region}
}
\tablewidth{0pt}
\tablehead{
\colhead{Region} & \colhead{Matched source} & \colhead{Unmatched source} & \colhead{Source contribution} & \colhead{Reference} 
}
\startdata
 $|\ell| \leq30\degr$ and $|b| \leq15\degr$ & 33 & 860 & 20--25\% & \cite{bouchet_diffuse_2011} \\
 $|\ell| \leq60\degr$ and $|b| \leq10\degr$ & 40 & 1162 & 20--30\% & \cite{strong_diffuse_1996} \\
 $|\ell| \leq30\degr$ and $|b| \leq5\degr$ & 25 & 551 & 20--30\% & \cite{strong_comptel_1999} \\
 $|\ell| \leq45.7\degr$ and $|b| \leq45.7\degr$  & 46 & 1923 & 8--25\% & \cite{siegert_diffuse_2022} \\
\enddata
\tablecomments{Unmatched source number is the sum of \bat\ and \lat\ sources within the region, excluding the matched sources.}
\end{deluxetable*}

% Reference
%%%%%%%%%%%%%%%%%%%%%%%%%
%%%%%%%%%%%%%%%%%%%%%%%%%

%\bibliography{references}
\bibliography{output}



\end{document}